\documentclass[12pt]{article}
\usepackage{graphicx,psfrag}
\usepackage{epsfig}
\usepackage{amsmath}
 \usepackage{amscd}
 \usepackage{sidecap}
 \usepackage{wrapfig}
 \usepackage{amssymb}
\usepackage{epsf,float}

 \newread\testifexists
 \def\GetIfExists #1 {\immediate\openin\testifexists=#1
     \ifeof\testifexists\immediate\closein\testifexists\else
     \immediate\closein\testifexists\input #1\fi}

 \usepackage{gthstyle}\usepackage{amsfonts}
 \usepackage{amssymb}
 \immediate\openin\testifexists=e
     \ifeof\testifexists\immediate\closein\testifexists\else
     \immediate\closein\testifexists\input e\fipsf

 \def\Bbb#1{\setbox0=\hbox{$\tt #1$}  \copy0\kern-\wd0\kern .1em\copy0}

 \def\bbf#1{\setbox0=\hbox{$#1$} \kern-.025em\copy0\kern-\wd0
         \kern.05em\copy0\kern-\wd0 \kern-.025em\raise.0433em\box0}

 \immediate\openin\testifexists=a
     \ifeof\testifexists\immediate\closein\testifexists\else
     \immediate\closein\testifexists\input a\fimssym.def  

 \def\ffract#1#2{\raise .2 em\hbox{$\scriptstyle#1$}\kern-.3em/
                 \kern-.2em\lower .15 em \hbox{$\scriptstyle#2$}}

 \def\part#1#2{{\partial#1\over\partial#2}}

 \newcommand{\pg}{\psfrag}
\def\ptl{\partial}\def\beq{\begin{equation}}
\def\nn{\nonumber}
\def\eps{{\epsilon}}

 \newcommand{\crlb}[1]{\label{#1}\\[2pt]}
 \newcommand{\eela}[1]{\quad\hbox{\scriptsize{#1}}\label{#1}\end{eqnarray}}
 \newcommand{\eelb}[1]{\label{#1}\end{eqnarray}}
 
 \newcommand{\newsecb}[2]{\section{#1}\label{#2}\setcounter{equation}{0}}
 
 \newcommand{\nolabels} {\def\eel{\eelb} \def\crl{\crlb} \def\newsecl{\newsecb}}

    \newcommand\publishversion{\nolabels\setlength{\textheight}{9in}\setlength{\oddsidemargin}{.0in}
    \setlength{\textwidth}{6.4in}\setlength{\topmargin}{-0.4in}}

\publishversion
\begin{document}
 \begin{titlepage}

\title{\begin{center}\hfill {\small MZ-TH/10-25}   \vspace{0.3cm} \\
 \large\bf{COMPACT TIME AND DETERMINISM  FOR BOSONS:  
foundations} 
\end{center}}

\author{Donatello Dolce\thanks{e-mail: dolce@thep.physik.uni-mainz.de}\\
\smallskip \\
\textit{Institut f\"ur Physik (WA THEP)}\\
\textit{ Johannes-Gutenberg Universit\"at}\\
\textit{D-55099 Mainz}\\
\textit{ Germany}}

 \maketitle

\begin{abstract}
Free  bosonic fields are investigated at a classical level by imposing  their charac\-teristic de Broglie periodicities  as constraints. 
In analogy with finite temperature field theory and with  extra-dimensional field theories, 
this compac\-tification  naturally leads to  
a quantized energy spectrum. 
 As a consequence of the relation between periodicity and energy arising from the de Broglie relation, the compactification must be regarded as  dynamical and local. 
The theory, whose fundamental set-up is presented in this paper, turns out to be consistent with special relativity and in particular respects causality. 
The non trivial classical dynamics of these periodic fields show remarkable overlaps with ordinary quantum field theory. This can be interpreted as a generalization of the AdS/CFT correspondence.  
\end{abstract}
\vspace{4.0cm}



   \end{titlepage}

\eject

   \maketitle
  \pagebreak
\tableofcontents{}

\addcontentsline{toc}{section}{Introduction}
\section*{Introduction} \label{sec:0}

Time is a concept that has always played a central role in physics.   Its operative definition   is given by counting the number of cycles of a phenomenon ``supposed" to be periodic. 
The importance of the assumption of periodicity is also present in the A. Einstein's definition of a relativistic clock
 \begin{quote}
\textit{``By a clock we understand anything characterized by a phenomenon passing periodically through
identical phases so that we must assume, by the principle of sufficient reason, that all that
happens in a given period is identical with all that happens in an arbitrary period".}  ~~~~~~~~~~~~~~~~~~~~~~~~~~~~~~~~~~~~ A. Einstein \cite{Einstein:1910}
\end{quote} 
As the Einsteinian  concept of time itself, the ``period" 
considered in the above definition 
implicitly has a local and dynamic nature related to the motion of the reference frames. 
 A manifestation of this aspect is the relativistic Doppler effect, which is a direct consequence of the  time interval variations induced by the Lorentz transformations. 
 In fact, in modern physics time emerges from the Minkowski metric as a fourth dimension in addition to the spatial ones.

Despite the great success of  special relativity, among the most challenging  issues in modern physics  there are still those concerning  the notion of time.
The relativistic laws are compatible with an inversion of the time arrow. 
On the other hand, from a statistical point of view, the time arrow is  fixed by the second law of thermodynamics, which states that the total entropy of the universe must increase for probabilistic reasons.  
Also in non-relativistic quantum mechanics, time plays a peculiar role with respect to the spatial variables. 
Within the Hamiltonian formulation, the time dimension  emerges dynamically through the Schr\"odinger equation, whereas  the Lagrangian formulation highlights a connection between statistic and quantum mechanics. 
In fact there is a close analogy between the  Boltzmann formulation  of  statistical mechanics and  the Feynman   path integral formulation of quantum mechanics \cite{Zeh:1992vf}.
It is well known that the quantization of  three-dimensional statistical systems is achieved  by adding  a periodic time dimension of Euclidean type \cite{Zee:2003mt,Zinn-Justin:2000dr,Kapusta:1989tk}. 
This is one of the basic assumptions of field theory at finite temperature. 
The statistical quantum systems  studied in this way are those of  quantum fields  at  thermal equilibrium, whose  Euclidean time periodicity is  proportional to the inverse  of the  temperature.

Theories with  time periodicity are studied in different  branches of physics such as  quantum mechanics,  thermodynamics,  statistical mechanics,  cosmology and  elementary particle physics. 
 In the canonical formulation of special relativity the space-time dimensions are implicitly assumed to be infinite. 
 However, recently the idea of a compact time dimension at a cosmological scale, \textit{i.e.} a universe intrinsically periodic or cyclic, has become of growing interest for the understanding  of  the origin  and   end of the universe \cite{Penrose:2008,Steinhardt:2002kw,Nielsen:2006vc,Nielsen:2006th}.  
 Moreover, in  black hole thermodynamics, the space-time curvature due to the black-hole mass $M_{\odot}$ yields  a metric which, upon Wick rotation, gives a time  periodicity  $8 \pi G M_{\odot} $ and thus the Hawking radiation \cite{Zee:2003mt}.  Concerning general relativity, we only note that the Einstein equations admit  time periodic solutions \cite{Kong:2008eu}. 
We note that the Lorentz  transformations emerging from the  Minkowskian metric fix the differential structure of special relativity but they do not prescribe any particular restriction to the boundary conditions that must be imposed on the space-time dimensions. 
On the other hand, the solution of  relativistic differential systems requires the assumption of suitable  conditions  on the four-dimensional boundaries in which the theory is embedded. 
Generalizing the Hamilton principle, the important requirement is that these  conditions minimize the relativistic action on the boundaries, \textit{i.e} they are Hamiltonian constraints 
\cite{Henneaux:1998ch}. 
For this reason the fields  are usually assumed  to have  fixed values at initial and final times, however others typologies of boundary conditions  such as periodic or antiperiodic ones are allowed  \cite{Csaki:2003dt}.  

 First we restrict our study to elementary isolated systems, 
  where the Bohr-Sommer\-feld quantization condition says that, in a give potential, the  allowed phase-space orbits are those with an integer number of periods. 
 This is a periodicity condition. 
 A typical application of this approach are the Bohr orbitals in the Hydrogen atom. Historically this was one of the first evidences of quantum mechanics. 

Continuing using  as simple as possible arguments, we  note that a naive way to obtain a quantization of the energy is to set its conjugate variable, namely the physical time, on a finite interval. 
This is in close  analogy with the Matsubara theory  \cite{Matsubara:1955ws} where the assumption of a periodicity condition in the Euclidean time 
yields  a discretized thermal energy tower whose levels are known as Matsubara energies. 
 Another similitude is given by the Kaluza-Klein theory \cite{Kaluza:1921tu,Klein:1926tv} where a  discretized mass spectrum is obtained  by imposing that the field is embedded in a compact  spatial extra-dimension. Since the proper time is the conjugate variable of the invariant mass, we will address it as ``virtual" extra dimension.
 Provided the identification of the extra coordinate with the time is made, in extra dimensional theories the determination of the mass spectrum is obtained from a differential system   analogous  to the Schr\"odinger one \cite{Dvali:2001gm,Carena:2002me,Karch:2006pv}. 
 This is a consequence  of the fact that the Klein-Gordon equation is the relativistic generalization of the   Schr\"odinger equation.

  The key assumption of this work is a generalization of 
  the de Broglie hypothesis \cite{Broglie:1924,Broglie:1925}:  
every  field has a given angular frequency $ \bar \omega $
(as long as it does not interact); the energy of the related quanta  $\bar E = \hbar \bar \omega = h / T_t $  is fixed by the inverse of the period $T_t $ through the Planck constant $h$. 
 We will impose such de Broglie periodicities $T_t$ as  constraints to the fields  
and we will obtain a quantized tower of energy resonances with  gap $\bar E = h   / T_{t}$. 
 The energy resonances  associated to such a $T_t$ periodic field will be interpreted in  terms of  quanta with energy $\bar E$. 
 To obtain a consistent relativistic invariant theory we will of course also consider  the periodicities induced by the time dimension on the modulo of the spatial dimensions and, for massive fields, on the proper time.
 To see this consistence it is important to bear in mind that 
  we will always impose the usual de Broglie space-time periodicities  of the relativistic fields as constraints. 
  
  To generalize the ``old" quantum theory we will focus on periodic scalar fields, \textit{i.e.}   packets of free relativistic waves satisfying the same periodic boundary conditions. 
 Such a harmonic system, similar to acoustic waves, is one of the simplest and most fundamental systems in nature. 
  By  considering compact space-time dimensions, the periodic fields will be described by a quantized energy-momentum spectrum. 
 The main difference with to Kaluza-Klein theory is that the compactification periodicities  are now fixed by  dynamical parameters like the energy or the momentum and not  by an invariant parameter like the mass. Therefore such a  compactification must be regarded as dynamical and local, and not statical and invariant as in the Kaluza-Klein model. 
In this way, sec.(\ref{quantum}), we will find for both massless and massive fields that every energy eigenmode has the correct relativistic  dispersion relation, so that the quantization that we obtain corresponds  to the normal ordered energy spectrum of the ordinary quantum relativistic fields. 
 In particular the mass is given by the inverse of the proper time (``virtual" extra-dimension) periodicity which in turn fixes the upper limit of the physical time periodicity through Lorentz transformations. 
 This intrinsic time periodicity is know as de Broglie periodic phenomenon or de Broglie internal clock of massive particles. 
  A hypothetic boson with the mass of an electron has an intrinsic rest periodicity, proportional to the Compton wave length, of about $10^{-20} s$. 
Even for a  mass as light as that, the periodic dynamics are extremely fast.    
For instance,  the  oscillation period of the  Cs-$133$ atom, which is used in the operative definition of time, is of the order of $10^{-10} s$. 
Remarkably,  this intrinsic periodicity of massive particles  has been indirectly observed only in a recent experiment \cite{2008FoPh...38..659C} for electrons (which for the scope of this paper  can be though of as  fields with antiperiodicity $T_t$).
 
The theory is based upon relativistic waves with the boundary conditions that minimize the relativistic action, thus every perturbation in a given point propagates with the retarded and advanced potential - as well as the information. 
The resulting periodicities for these fields are indeed dynamical and local.
This means that the compactification radius must \textit{not} be regarded as static since it changes according to the relativistic causality and to energy conservation. 
In other words,  interactions destroy the original periodicity so that the system passes from a periodic regime to another periodic regime depending on the amount  of  energy exchanged, just as in  Compton scattering. 
 In this way it is possible to order events in time. We will conclude that the dynamical compactification arising from our theory respects all the fundamental requirements for a well formulated  notion of relativistic time. 

To have a simple image of our assumption one should remember how acoustic waves are described in terms of objects vibrating within compact spatial dimensions. 
In a full relativistic generalization of acoustic waves, our  quantum-relativistic fields can be regarded as imbedded in compact space-time dimensions. 
Thus  we  want to consider all the harmonics modes allowed by the de Broglie periodicities, not only the fundamental ones of the usual approach.  
Ordinary field theory, which is supposed to describe every elementary systems (or systems that appear to be elementary in a given approximation), is based upon de Broglie waves whose characteristic periods can be regarded as internal (de Broglie) clocks. 
The usual relativistic time axis is defined by reference to the ``ticks" of these  periodic phenomena, in particular to the ones of the Cs-$133$ atomic clock. 
For  massless (electromagnetic or gravitational) fields these periodicities can be in principle infinite.  In general the periodicities vary through energy exchange and the combination of two periodic systems results in ergodic (not exactly periodic) evolutions. 
Every value of the time axis can be characterized by a combination of the different phases of these de Broglie internal clocks of the elementary fields considered. Hence the external time axis can be dropped. 
Considering the Einstein's definition of relativistic clock we notice that the physical information for the fields is in the single periods. 
This leads to dynamic compact intervals with periodic conditions. 
  
The aim of this investigation is to stress the analogy between  dynamic periodic fields  and  the usual  quantum fields, sec.(\ref{Periodic:dynamics}).  
 From the analogy with the Kaluza-Klein theory we will find that the energy eigenstates  has  a Marcovian time evolution which is described by the  Schr\"odinger equation.  
 Being stationary waves, they  form a complete set  with an underlying inner product which can be used to build a Hilbert space. 
Formally, with this at hand,  the Feynman path integral  for free bosonic fields arises  without any further assumptions.  
Due to the periodic nature of the fields we will be able to extract the commutation relations as well as other aspects  of quantum mechanics such as 
 the Heisenberg relation and a generalization of the Bohr-Sommerfeld condition. 
   
Since quantum behaviors arise from a classical system  we can talk about determinism or pre-quantization.    A proposal for the possible deterministic nature of quantum theory has been given by 't Hooft
who  \cite{'tHooft:2001ar,'tHooft:2001fb,'tHooft:2001ct} had shown that there is a close relation between a classical particle moving on a circle and the quantum harmonic oscillator.  This model can be thought of as having a  periodic time  on a lattice, \cite{Arkani-Hamed:2001ca,Berezhiani:2002et}. 
Similarly to a rolling dice, if the periodicity  is  too fast,  at every observation  the system  results in a different phase of an apparently aleatoric evolution \cite{Elze:2002eg}. 
The system can only be described by using a transition probability from one state to another which turns out to be in agreement with the usual quantum rules for the harmonic oscillator.  
Within these scenarios important efforts to formulate quantum mechanics from a classical theory with compact  extra-dimensions  have been made in \cite{Elze:2003nu,Elze:2003ws,Elze:2003tb}. 
In this case ergodic dynamics reproduce  quantum behaviors in terms of an emerging effective time.
It is important to note  that, in our theory, by supposing that the quantum behavior arises from periodicity boundary conditions, we are   avoiding the  introduction of hidden variables  and at the same time we are implicitly introducing  a non-locality, so that our model is not constrained by Bell's theorem \cite{Nikolic:2006az}.

\section{Dynamic approach to compactified time}\label{quantum}

The differential structure of relativistic kinematics is based on the four-dimensional  Minkow\-ski metric 
\begin{equation}\label{mink:metr:4d}
d s^{2}=c^2 d\tau^2 =  c^2 dt^2 - d\mathbf{x}^2~,
\end{equation}  
and the related Lorentz transformations.
In this metric a Klein-Gordon complex  field $\Phi_{KG}(\mathbf{x}, t) $ with mass $M$ obeys  the equation  
$
(\ptl_{\mu}\ptl^{\mu} + {  M^{2} c^{2}}/{ \hbar^{2} }) \Phi_{KG}(\mathbf{x}, t) = 0 
$ 
 and it holds the relativistic dispersion relation
$ 
E^{2}(\mathbf{p}) = |\mathbf{p}|^2 c^{2}+  M^2 c^{4} 
$. 
It is worth noting  that  the solutions of this differential system depend on the boundary conditions imposed on the field and that neither the  Minkowskian  metric nor the Lorentz transformations  prescribe restrictions to them.  
In general, boundary conditions must be chosen such as to be consistent with the variational principle applied on the boundaries \cite{Henneaux:1998ch}.  
For simplicity in this preliminary discussion we will concern only with time dimension boundaries. 
 Thus, for a generic scalar field  $\Phi(\mathbf{x}, t)$ with time evolution inside the an interval $t \in [t', t' + 2 \pi R_{t}]$ and no additional boundary terms, the important requirement is that  
\begin{equation}\label{bound:act:var}
\int d^3 x \left[ \delta \Phi(\mathbf x,t) \partial_{t} \Phi(\mathbf x, t) \right]_{t'}^{t'+ 2 \pi R_{t}} \equiv 0~.
\end{equation}
In  ordinary field theory this relation is satisfied by choosing  fixed values of the field at the initial and final times $\delta \Phi(\mathbf x,t') = \delta \Phi(\mathbf x,t' + 2 \pi R_{t}) \equiv 0$.\footnote{The condition eq.(\ref{bound:act:var}) is invariant under time translations.} 
 However also  periodic,  antiperiodic  or (more generally) combinations of Dirichlet ($ \Phi = 0$) and Neumann ($\partial_{t} \Phi = 0$) boundary conditions are compatible with the variational principle  
 \cite{Csaki:2003dt}.
  From eq.(\ref{bound:act:var}) we see  that  these  conditions  have the same formal validity of the usual  conditions  assumed in ordinary relativistic field theory - they act as Hamiltonian constraints. 
  
  In particular, we want to explore at a classical level the physics of a free scalar  field $\Phi(\mathbf x, t)$  
 imposing periodicity as a constraint,  which means the following condition 
 \begin{equation}\label{peridic:BC}
\Phi(\mathbf x, t') \equiv \Phi(\mathbf x, t' + 2 \pi R_{t})~. 
\end{equation}
Imposing  periodicity along the physical time dimension in order to satisfy eq.(\ref{bound:act:var}) and to fix a particular solution of  the field equation, doesn't necessarily mean to localize the field $\Phi(\mathbf x, t)$ in a particular space-time region.
Skipping mathematical details, an equivalent way to interpret the condition  eq.(\ref{peridic:BC}) is to take the time either on a compact interval $t \in [t',t'+2 \pi R_{t}]$  or on the whole interval in $ \mathbb{R}$ where periodic condition eq.(\ref{peridic:BC}) is supposed to be satisfied. In both cases the whole physical information is contained in a single period and we can restrict our analysis to this region. 
Using a terminology common in Kaluza-Klein theories we  can say that  the time dimension is compactified on a circle $t \in \mathbb{{S}}_{R_t}^{1}$ with compactification radius $R_{t}$.\footnote{This does not  mean to deform the flat Minkowskian background eq.(\ref{mink:metr:4d}).} 

Provided analogous periodic conditions along the spatial and, for massive fields along the proper time dimensions, such as to guarantee covariance, we shall see that this theory of periodic fields is consistent with special relativity.  
 In other words we want to impose the natural (de Broglie) periodicities of the relativistic fields as a constraints to determine the solution of the Klein-Gordon differential equation in every space-time point.   In this way we want to generalize the ``old" formulation of quantum mechanics: free bosonic fields are supposed to have intrinsic periodicities $T_t$, so that the energy of the related quanta $\bar E = \hbar \bar \omega = h / T_t$ depends on the inverse of the time period $T_t$. This assumption can be regarded as the combination of the Newton's law of inertia with the de Broglie hypothesis. 
 
\subsection{Massless bosonic fields}\label{Massless:bosons}

Relativistic massless fields with  time periodicity $T_{t}$ imposed as a constraint   
are described by the following massless Klein-Gordon action  
\begin{equation}\label{KG:act:massless}
{\mathcal S}[T_{t}] = \frac{1}{2} \int_{0}^{\lambda_{x}}  d^3x \int_{t'}^{t'+T_{t}} dt  \left[ \partial_\mu \Phi^{*}(\mathbf x,t) \partial^\mu \Phi(\mathbf x,t) \right] ~. 
\end{equation}
To satisfy the variational principle on the time boundaries, eq.(\ref{bound:act:var}), we impose  the periodic condition eq.(\ref{peridic:BC}).  
Thus we can  decompose the field in  frequency eigenmodes
\begin{equation}
\Phi(\mathbf{x},t) = \sum_{n=-\infty}^\infty   \Phi_{n}(\mathbf{x}) u_{n}(t) ~,
\end{equation}
where the time evolutions are given by 
$ 
 u_{n}(t) = \exp[{- i \omega_{n} t }]
$ 
and the  angular frequency eigenvalues must be  
$ 
\omega_{n} = {n}/{R_{t}}
$. They are in fact the harmonic modes of a string with periodic boundary conditions.   
The eigenfunctions $u_{n}(t)$ form a complete and orthogonal set, so that we can 
decompactify the action along the time dimension 
\begin{eqnarray}\label{time:decompact}
{\mathcal S}[T_{t}] 
 &=& \frac{T_{t}}{2}  \int_{0}^{\lambda_{x}}  d^3x \sum_{n}  \left[ \ptl_{i} \Phi^{*}_{n}(\mathbf x)\ptl^{i} \Phi_{n}(\mathbf x)  + \frac{ \omega_n^{2}}{c^{2}}  | \Phi_{n}(\mathbf x)|^{2} \right] ~, 
\end{eqnarray}
obtaining a sum over three-dimensional actions of  the  eigenmodes $\Phi_{n}(\mathbf x)$.

Assuming the de Broglie relation, we proceed similarly to the Kaluza-Klein theory or to the Matzubara  theory and we associate  the quantized frequency spectrum $\omega_{n}$ of the $n$-th eigenmode to quantized energy spectrum $E_{n}$.   The proportionality  constant is the reduced Planck constant $\hbar$. 
In fact, 
calling  the wave number for the $n$-th  eigenmode $\mathbf k_{n}$, we find
\begin{equation}\label{onshel:massless:wave}
\square \Phi(x,t) \equiv 0 \rightarrow   \left( \frac{n}{R_{t}}\right)^2 - \mathbf{k}^2_{n} c^2 = 0~.
\end{equation}
Comparing with  the massless dispersion relation
$
 E^2 - |\mathbf{p}|^{2} c^2 = 0
$, 
it is natural to assume  
\begin{equation}\label{discr:ener:lev}
E_{n}  \equiv \hbar \omega_{n} =   \frac{n \hbar}{R_t}~.
\end{equation}  
After the decompactification  we have a tower of energy eigenstates exactly  as in  extra-dimensional theories one finds a tower of massive eigenstates.
 The  main difference with extra-dimensional theories is  that the mass  (and thus the compactification radius of the extra-dimension) is four-dimensional invariant (but not five-dimensional invariant) whereas the energy of the field is a dynamical quantity. 
 As we will discuss in details, the time compactification radius must be therefore regarded as dynamical. 
 Here it is worth to note that every mode has a positive defined energy, since Kaluza-Klein modes have always positive defined masses (no tachyonic modes). 
 
 To obtain a consistent relativistic theory we must consider also the compactification of the spatial dimensions. In particular, since we are assuming  massless on-shell fields, we find that the time periodicity  induces a periodicity  on the absolute values of the spatial dimensions as well. From  eq.(\ref{onshel:massless:wave}) it follows that  the absolute values of the momenta must be as well discretized
\begin{equation}\label{discr:mom:lev}
|\mathbf p_{n}| =  \hbar |\mathbf k_{n}| =  \frac{n \hbar }{R_{t} c}~ .
\end{equation} 
In other words, since in the massless case the field is on the light-cone $d s^{2} = 0$, we get   $c^2 dt^2 = d\mathbf{x}^2$  and thus an induced spatial periodicity 
\begin{equation}\label{massless:spat:period}
\lambda_{x} =  T_t c = \frac{n h}{ |\mathbf p_{n}|} ~.
\end{equation} 

The gap between the energy levels can be expressed in terms of the energy of the fundamental level (the energy of the  eigenmode with $n = 1$). 
We denote the quantities related to this fundamental level  with the bar sign, so it has energy, momentum  and angular frequency   $\bar E$, $\mathbf{\bar p}$  and $\bar \omega$ respectively. Hence,  eq.(\ref{discr:ener:lev}), the compactification radius is fixed by the fundamental energy through the following relation 
\begin{equation}\label{fund:level}
R_{t}(\mathbf{\bar p}) \equiv \frac{\hbar }{\bar E (\mathbf{\bar p})} = \frac{1 }{\bar \omega (\mathbf{\bar p})}~.
\end{equation}

This relation (namely the de Broglie relation)  emerges naturally from the periodic field formulation and we take it as one of the   basic assumption of this work.  In particular the energy has a geometrical interpretation in terms of the compactification length (compression) of the vibrating string.  
The dispersion relation for the first eigenmode in this massless case, eq.(\ref{discr:mom:lev}),  is 
\begin{equation}\label{massless:disp:relat}
\bar \omega (\mathbf{\bar p}) = \frac{|\mathbf{\bar p}| c}{\hbar} ~.
\end{equation}

The relevant aspect of this result is that there is a  discretized (quantized) energy spectrum and the Planck constant $h$ relates the temporal period to the inverse of the energy.
From these considerations we finally check that the four-momentum of the fundamental level and  the space-time compactification radiuses can be written respectively as  $\bar p_\mu = (\bar E/ c, \mathbf{\bar p})$ and  $R_\mu = (c R_t, \mathbf R_x)$, where $|\mathbf R_x| = R_x 
= \hbar / |\mathbf {\bar p}|$.  Generalizing the de Broglie hypothesis, the  fundamental  compactification conditions eq.(\ref{fund:level}) and eq.(\ref{massless:spat:period}) can be written with the following notation
\begin{equation}\label{fund:level:covar}
R_\mu = \frac{\hbar}{\bar p^\mu }~.  
\end{equation}

\begin{figure}{
\pg{E}{$ E_{n}(\mathbf p)$}
\pg{R}{$ R_{t}(\mathbf p)$}
\pg{0.5}{}
\pg{1}{}
\pg{w}{$\hbar  \bar \omega(\mathbf{\bar p})$}
\pg{p}{$\mathbf {\bar p}$}
\pg{n}{$n \hbar  \bar \omega(\mathbf{\bar p})$}
\hspace{-0cm} a)~~~\epsfig{file=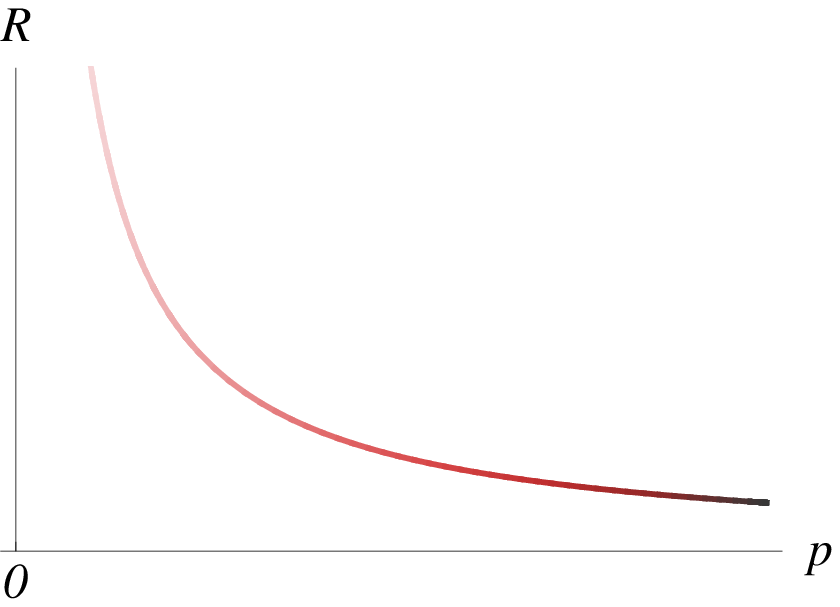,scale=0.77} ~~~
b)~\epsfig{file=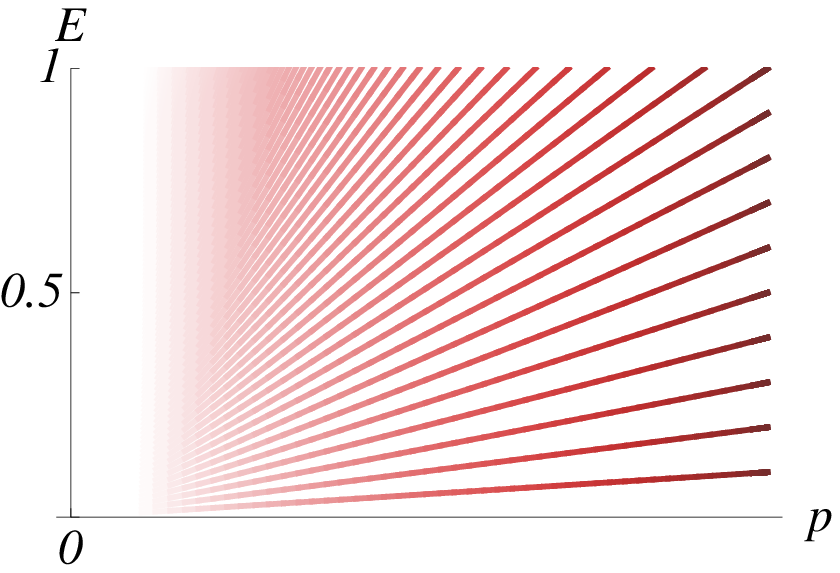,scale=0.77} 
 \caption{Spectral behavior for a massless periodic field   as a function of  the fundamental momentum $\mathbf{\bar p}$. 
 Fig.(a)  shows the variation of the compactification radius  $R_{t}(\mathbf{\bar p}) \equiv {\hbar }/{\bar E (\mathbf{\bar p})}$, according to $\bar E(\mathbf{\bar p}) = \hbar \bar \omega (\mathbf{\bar p}) = {|\mathbf{\bar p}| c}$.  
 Fig.(b) shows the massless relativistic dispersion relation of the resulting energy spectrum $E_{n} (\mathbf{\bar p}) = n \hbar \bar \omega (\mathbf{\bar p})$. 
 In the limit of  zero momentum the fundamental compactification radius tends to infinity  giving a continuos energy spectrum.}
\label{dispers-massless}}\end{figure}

From eq.(\ref{fund:level}) and eq.(\ref{massless:disp:relat}) we see that the usual relativistic  massless field  is obtained in the limit of infinite compactification radius, or equivalently by taking  the radius constant and  doing the limit of small $\hbar$, see fig(\ref{dispers-massless}.a). 
Both  limits tend to  decrease  the gap between the energy levels, thus obtaining  a continuous energy spectrum, as shown on the left side of fig(\ref{dispers-massless}.b). 
On the contrary, in the limit of large $\hbar$ or high fundamental frequency $\bar  \omega (\mathbf{\bar p})$, we obtain a well discretized energy spectrum, right side of fig(\ref{dispers-massless}.b). 
For the energy levels of  periodic fields at the thermal equilibrium it is natural to assume the Boltzmann occupation probability $\propto \exp[- n \bar E / K \mathcal{T}]$ , where $\mathcal{T}$ is the temperature  of the thermal bath.  As explained in details in  the version 4 of this paper \cite{Dolce:2009ce},  if $ K \mathcal{T} \gg \hbar \bar \omega$  many energy levels are  populated and the field can be approximated by a continuos energy spectrum.
 On the other hand, if $ K \mathcal{T} \ll \hbar \bar \omega$, only few levels are populated; here the quantization of the energy spectrum is manifest. This is the condition needed to avoid the UV catastrophe in  the black body radiation or to describe the single photon limit. 
 
\subsection{Massive bosonic fields}\label{Massive:bosons}

The key assumption for  a massive relativistic field is that it is possible to choose a reference system (the rest frame) where the real time and the proper time can be identified.\footnote{For sake of simplicity we address:  the invariant time as the proper time,  the euclidian time  as the imaginary time, and  the  Minkowskian time as the real or physical time.} 
Therefore, for massive fields, we must consider that the compactification of the real time induces a compactification of the proper time, as well.
 
\begin{figure}{
\pg{l}{$R_{\tau} = {\hbar}/{\bar M c^2}$}
\pg{E}{$ E_{n}(\mathbf p)$}
\pg{R}{$ R_{t}(\mathbf p)$}
\pg{0.5}{}
\pg{m}{\!\!\!\!\!\!\!\!\!\!${\bar M} c^{2}$}
\pg{k}{\!\!\!\!\!\!\!\!\!\!\!\!\!\!$n \bar M c^{2}$}
\pg{w}{$\hbar  \bar \omega(\mathbf{\bar p})$}
\pg{p}{$\mathbf {\bar p}$}
\pg{n}{$n \hbar  \bar \omega(\mathbf{\bar p})$}
\hspace{-0cm}
a)~~~\epsfig{file=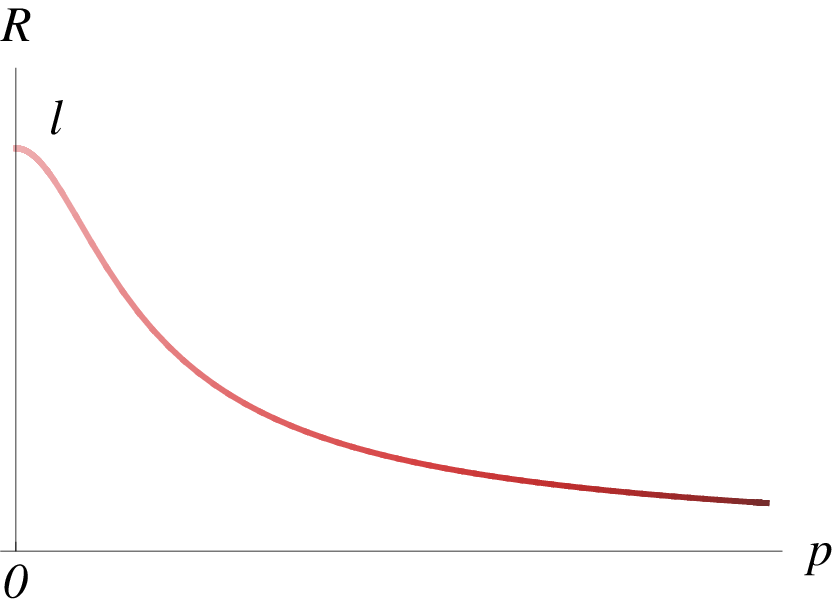,scale=0.77} ~~~
b)~\epsfig{file=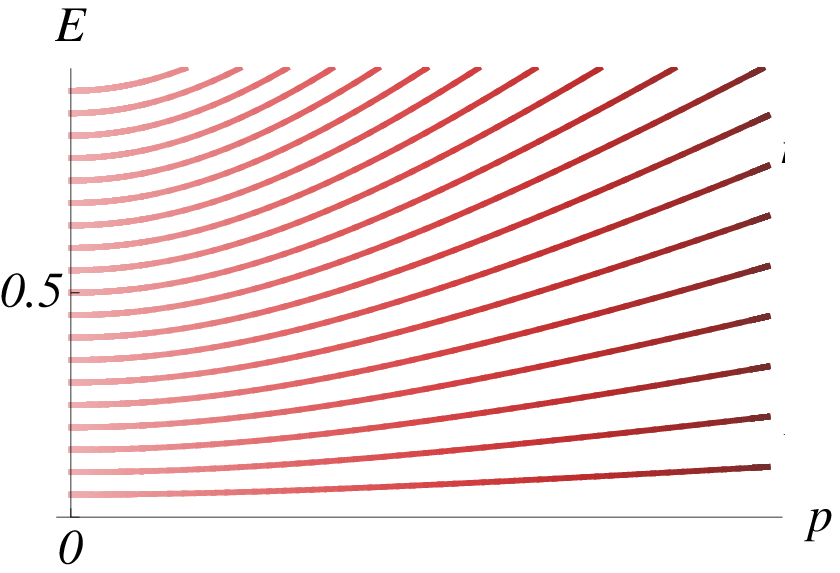,scale=0.77} 
\caption{Spectral behavior for a massive periodic field with 
mass $\bar M = \hbar / c^2 R_\tau$  as a function of  the fundamental momentum $\mathbf{\bar p}$. 
Fig.(a) shows the  variation of the compactification radius $R_{t}(\mathbf{\bar p}) \equiv {\hbar }/{\bar E (\mathbf{\bar p})}$, according to $\bar E(\mathbf{\bar p})=\hbar \bar \omega (\mathbf{\bar p}) = (\mathbf {\bar p}^{2} c^{2} + \bar M^{2} c^{4} )^{1/2}$.  
Fig.(b) shows the massive relativistic dispersion relation of the resulting energy spectrum  $E_{n} (\mathbf{\bar p}) = n \hbar \bar \omega (\mathbf{\bar p})$.  
The proper time compactification radius $R_{\tau}$ fixes the upper limit for $ R_{t}(\mathbf{\bar p})$. 
}
\label{dispers-massive}}\end{figure}

We approach the theory as a  Kaluza-Klein theory for a massless five-dimensional  Klein-Gordon field with periodic extra-dimension $s$ and periodic real time.  
In fact the resulting  five-dimensional  metric is    
$
 dS^2 =  c^2 dt^2 - d\mathbf{x}^2 -  ds^2 \equiv 0
$,  
so that assuming $s = c \tau $  we recover the usual four-dimensional  Minkowskian metric eq.(\ref{mink:metr:4d}). 
For this reason we will say that the proper  time $\tau$ acts as a ``virtual" extra-dimension\footnote{The  Kaluza-Klein  quantized mass spectrum  can be regarded as a consequence of the fact that the extra-dimension acts similarly to a proper time. In fact, in field theory the conjugate variable of the proper time is the mass and therefore, by putting the proper time in a compact segment we have a quantization of the mass spectrum, that is of the rest frame energy spectrum. } whose  length  is therefore fixed by the time periodicity in the rest frame. 
We temporarily write the scalar field as a double sum over  eigenstates, one over discrete energies because of the periodic time and one over discrete mass eigenmodes because of the induced periodicity on the proper time
\begin{equation}
\Phi(\mathbf x, t , s) 
= \sum_{n_{t},n_{s}} e^{-i n_{t} \bar \omega t + i n_{s} \bar \sigma s} \Phi_{n_{t}, n_{s}} (\mathbf x)~,
\end{equation} 
where $\bar \sigma = {2 \pi}/{ \lambda_{s}}$ and $\lambda_{s}= c T_{\tau}$. 
The (virtual) five-dimensional Klein-Gordon action  is 
\begin{eqnarray}\label{KG:M:5D}
{\mathcal S}[T_{t}]   &= &   \frac{1}{2  \lambda_{s}} \int^{ \lambda_{s}}_{0} d s \int^{ T_{t}}_{0} d t  \int_{0}^{\lambda_{x}}  d^{3} x   \left[  \partial_M \Phi^{*}(\mathbf x, t , s)\partial^M \Phi(\mathbf x, t , s)  \right] ~. 
\end{eqnarray} 
Decompactifying the proper time, in analogy with eq.(\ref{time:decompact}) and eq.(\ref{fund:level}), we have obtained a tower of  (virtual) four-dimensional Kaluza-Klein fields  with  invariant mass gap 
\begin{equation}\label{mass:gap:def}
\bar M c^{2} \equiv c \hbar \bar \sigma = \frac{\hbar }{ R_{\tau} }~. 
\end{equation}
Then we decompactify the real time obtaining a double tower of three-dimensional eigenmodes $ \Phi_{n_{s}, n_{t}}(\mathbf x)$ which satisfy the dispersion relation
\begin{equation}\label{doublesum:disp:rel}
n^{2}_{t} \frac{ \bar \omega^{2}}{c^{2}} =  \mathbf k_{n_{s}, n_{t}}^{2} + n_{s}^{2} \frac{\bar M^{2} c^{2}}{\hbar^{2}}  ~.
\end{equation}
 As it is evident at constant spatial separation ($d\mathbf{x}^2=0$) or equivalently at zero momentum ($\mathbf k_{n_{s}, n_{t}} \rightarrow 0$), 
where the proper time and real time periodicities ca be identified ($d\tau^2 =   dt^2$),  
we obtain the condition $n_{s} = n_{t} = n$. In fact, there is a single periodicity which is induced to the other dimensions and the final result must be a single sum over eigenmodes. 
 Finally from eq.(\ref{doublesum:disp:rel}) we  obtain 
\begin{equation}\label{massive:disp:relat}
  \bar \omega(\mathbf {\bar p} ) =  \frac{\sqrt{\mathbf {\bar p}^{2} c^{2} + \bar M^{2} c^{4} }}{\hbar}  ~,
\end{equation}
where $\hbar \mathbf k_{n_{s}, n_{t}} = n \mathbf{ \bar p}$. Thus
$ 
E_{n}(\mathbf {\bar p}) = n \hbar \bar \omega(\mathbf {\bar p}) =  n  \sqrt{\mathbf {\bar p}^{2} c^{2} +  \bar M^{2 }c^{4}}  ~.
$ 
As in the massless case eq.(\ref{massless:spat:period}), the induced spatial periodicity is $\lambda_{x} = h / | \mathbf {\bar p} |$.  The action eq.(\ref{KG:M:5D}) can be seen as a sum over energy eigenmodes with masses $M_{n} = n \bar M$, similarly to eq.(\ref{time:decompact}),\footnote{Note that the periodic massive field obtained in this section is not properly the Klein-Gordon field. The mass term arises by compactification.} \footnote{Extending our terminology we could say that the energy eigenmodes of the towers are ``virtual" Kaluza-Klein particle. }
\begin{equation}\label{one:sum:act}
{\mathcal S}[T_{t}]   =    \frac{  T_{t}}{2}   \int_{0}^{\lambda_{x}} d^{3} x \sum_{n}   \left[   n^{2} \frac{\bar \omega^{2}}{c^{2}} |\Phi_{n}(\mathbf x)|^{2}  - |\partial_i \Phi_{n}(\mathbf x)|^{2} - n^{2}  \frac{\bar M^{2} c^{2}}{\hbar^{2}} |\Phi_{n}(\mathbf x)|^{2}  \right] ~.
\end{equation}
We have  indeed obtained that the  discretized energy spectrum in terms of the compactification radius is still given by eq.(\ref{fund:level}), but now the eigenstates obey the dispersion relation of  relativistic massive particles, as shown in fig(\ref{dispers-massive}.b).  We note that this quantization is exactly the same one obtained from the usual normal ordered second quantization.  

From eq.(\ref{mass:gap:def}) we can interpret the mass as the inverse of the proper time periodicity: the bigger  the mass the shorter the proper time period. 
Since the energy is bounded from below by the mass $\bar E(\mathbf {\bar p}) \geqslant \bar M c^{2}$, the time compactification radius has the upper invariant bound $R_{t} (\mathbf {\bar p}) \leqslant R_{t}(0) = R_{\tau}$.  
Roughly speaking we can actually say that the mass fixes the inertia of the  motion, fig(\ref{dispers-massive}.a).
This proper time periodicity of the  field,  known as the de Broglie periodic phenomenon or the de Broglie internal clock of massive particles \cite{Broglie:1924,Broglie:1925,1996FoPhL}, is
\begin{equation}\label{prop:time:period}
T_{\tau} \equiv T_{t}(0)= \frac{h}{\bar M c^{2}} ~.
\end{equation}
 To this periodicity  the  invariant length 
\begin{equation}\label{compton:length}
  \lambda_{s} \equiv T_{\tau} c = \frac{h}{\bar M c}~
\end{equation}
  is  associated, which is nothing else than  the Compton wave length.      
A hypothetical light boson for instance with the  mass of an electron  has a Compton wave length $  \lambda_{s}  \sim 2 \pi \times 4 \cdot 10^{-13} m$ which leads to  the proper time periodicity   $T_{\tau} \sim   8 \cdot 10^{-21} s$. 
Even for such a light particle, this microscopic time scale can not yet be observed directly in the modern experiments.\footnote{It is worthwhile noting that the operative definition of time  is  given concretely by counting the number of periods of a ``well known periodic system''. 
 The most accurate definition of second is  the duration of 9,192,631,770 periods of the radiation corresponding to the transition between the two hyperfine levels of the ground state of the cesium 133 atom [SI].  By definition this period is $\sim 10^{-10} s$, whereas the best experimental resolution on resolving time known to the author is about  $\sim 10^{-16} s$ \cite{whurt:2005}. On the other hand, the periodicity explored at the $TeV$ scale is of the order of $\sim 10^{-27} s$.} However, as shown in a recent experiment \cite{2008FoPh...38..659C}, the modern techniques are reaching a sufficient precision to allow  indirect  evidences of the de Broglie internal clock of the electron - the proper time periodicity of eq.(\ref{prop:time:period}) must be regarded as a general property for massive particles. 

A massive periodic field turns out to be localized inside the Compton wavelength. In fact \cite{Dolce:2009ce}, the non-relativistic limit corresponds to a low intensity $|\mathbf {\bar p}| \ll \bar M c$ massive field where only the first energy level is largely populated. In this way we obtain  the usual non-relativistic free particle distribution (modulo the de broglie internal clock) $\phi(\mathbf x) \sim \exp{[- i \frac{\bar M c^2}{\hbar}t + i \frac{\bar M}{ \hbar}\frac{\mathbf x^2}{2 t}]}$.
This gives a consistent  interpretation of the dualism between waves and particles and also of the double slit experiment \cite{Dolce:2009ce}. 

\subsection{Lorentz transformations and covariance}\label{Lorentz:transformation}

To see that the periodicity in physical time and in  proper time are consistent with special relativity, we perform a Lorentz transformation $R_\tau = \gamma (R_t - \mathbf v \cdot    \mathbf R_x / c^2)$ with $\gamma =1/\sqrt{1 - \mathbf v^2/c^2}$, from the rest frame of the massive field to another reference frame at velocity $\mathbf v( \mathbf {\bar p})$. We find that \cite{1996FoPhL} the  relation 
$ 
\hbar = \bar M c^2 R_\tau 
 = E( \mathbf {\bar p}) R_t - \mathbf {\bar p}  \cdot  \mathbf R_x  
$ 
is  in perfect agreement with
 the behavior of the periodicity obtained in  eq.(\ref{massive:disp:relat}), see also fig.(\ref{dispers-massless}.a) and fig.(\ref{dispers-massive}.a).  
This shows why the time periodicity emerges as a dynamical constraint.   
It is different if observed from different reference systems, exactly as every other time interval in special relativity.  
Generalizing the notation of the massless case, $R_\mu$ is dynamically fixed by the four-momentum $\bar p_\mu$ as in the de Broglie hypothesis, see eq.(\ref{fund:level:covar}). 
In fact, the four-dimensional wavevector $k_\mu = (\bar \omega( \mathbf {\bar p})/c,\mathbf {\bar k})$ is Lorentz-covariant, whereas $\bar k_\mu R^\mu$,\footnote{Since $R_\mu$ is such that $\exp[-i \bar p^\mu x_\mu ] \equiv \exp[-i \bar p^\mu( x_\mu + 2 \pi R_\mu)] $, it turns out to be a contravariant four-vector.} being a  phase of the relativistic fields, is invariant under Lorentz transformations \cite{Broglie:1924,Broglie:1925,1996FoPhL}.

In the massive case the space-time compactification radiuses can be used to write the relativistic  dispersion relation  as $\bar M^2 c^4 = c^2 \bar p^\mu \bar p_\mu = (c \hbar/R_\mu)(c \hbar/R^\mu) $. Thus we find the constrain between the space and time periodicities 
$
{R^{-2}_{\tau}} = {R^{-2}_{t}(\mathbf {\bar p})} - {R^{-2}_{x}(\mathbf {\bar p})}
$, 
 so that we can associate to the mass $ \bar M$, to the energy $\bar E $ and to the momentum $ \bar {\mathbf p}$ a geometrical interpretation in terms of the compactification radius of the proper time $R^{2}_{\tau} $, of the real time $R^{2}_{t} $ and of the spatial coordinates $\mathbf{R}_x $, respectively. In few words, since the periodicities  that we are imposing in eq.(\ref{fund:level:covar}) are nothing else than the de Broglie periodicities, the model turns out to be automatically consistent with special relativity. 

At this point a small  digression  about   Lorentz invariance in theories with compact dimensions is in order. 
Through the decompactification of the time dimension in eq.(\ref{KG:act:massless}) we obtain an equivalent theory with an infinite sum of three-dimensional eigenmodes eq.(\ref{time:decompact}). 
The equivalence means that this infinite sum over three-dimensional modes is four-dimensional  Lorentz invariant as  the original formulation.  
  In general,  the Lorentz invariance breaking is not because of the compactness of a dimension but rather because that, in an effective lower dimensional theory, only a finite number of eigenmodes can be considered.  By generalizing the Higgsless gauge symmetry breaking mechanism induced by boundary conditions as in extra-dimensional theories  \cite{Csaki:2003dt} to a Yang-Mills theory with compact time, it is possible to show \cite{Dolce:2009ce} that there is a quantization of the magnetic flux and other  characteristic behaviors typical of a superconducting regime \cite{Weinberg:1996kr}. 

\subsection{Retarded potential and causality}\label{Delayed:potential}

Taking for simplicity only propagation of massless fields,  eq.(\ref{KG:act:massless}) is sufficient to fix the relativistic Green function.  
The  retarded or advanced Green function $G_{ret (adv)}(\mathbf x, t; \mathbf x', t')$ is formally the solution of the inhomogeneous relativistic wave equation with point-like source in $(\mathbf x', t')$. 
The Kirchhoff formulation allows us to write the solution for the field $\Phi(\mathbf x, t)$ as a source term plus  boundary terms\footnote{As in the introductory discussion about periodic boundary conditions, we are not considering explicitly the spatial boundary terms.} at generics initial and final time $t_{1}$ and $t_{2}$ \cite{Zeh:1992vf}
\begin{eqnarray}\label{delay:pot}
\!\!\!\Phi(\mathbf x , t) &=& \int_{t_{1}}^{t_{2}} d t' \int_{-\infty}^{\infty} d^{3 } x' G(\mathbf x, t; \mathbf x', t') j(\mathbf x', t') \nn \\
& -& \frac{1}{4 \pi c^{2}} \int_{-\infty}^{\infty} d^{3 } x' \left[ G(\mathbf x, t; \mathbf x', t') \ptl_{t'} \Phi(\mathbf x' , t') -  \Phi(\mathbf x' , t')
 \ptl_{t'} G(\mathbf x, t; \mathbf x', t') \right]_{t_{1}}^{t_{2}}\!.
 \end{eqnarray}
As we can  see from this equation,  a variation of the source term induces a retarded variation on the periodicity of the  field which must therefore be regarded as dynamical in the sense that it can vary through interactions.   In fact, only null sources $j(\mathbf x, t)$, or eventually sources with the same periodicity of the field, are compatible with  static space-time periodicities of the field itself.\footnote{This particular case leads to the so called billiard Green function or similar Green functions that must be not confused with the thermal Green function.} In the boundary terms the field acts similarly to a source term and the variation of periodicity of the field propagates in agreement with relativistic causality. This aspect is related to the dynamic and local nature of the compactification already discussed eq.(\ref{fund:level}) and can be interpreted in terms of the Huygens-Fresnel principle. 


At this point it is important to note that the theory is based upon relativistic waves. Thus the information  propagates in agreement with the relativistic causality.  
 By energy conservation,  a source term turned on in a given space-time point changes  the energy in another space-time point  after a  time delay, according to eq.(\ref{delay:pot}).  
 Therefore, assuming the dynamical compactification as in eq.(\ref{fund:level}) and  energy conservation, we observe that when the interaction is turned on, together with the energy irradiated, there is an induced variation of the compactification radius  from which more complicated and general time evolutions can be obtained.

  \begin{wrapfigure}{r}{0.46 \textwidth} 
 \begin{center}
\pg{t}{$t$}
\pg{w2}{$T^{\gamma}_{t}(\mathbf{\bar p}_{\gamma})$}
\pg{x}{$\mathbf x$}
\pg{w1}{$T^{pt}_{t}(0)$}
\pg{w3}{$T^{pt}_{t}(\mathbf{\bar p'})$}
\pg{c}{$d/c$}
\pg{d}{$d$}
\epsfig{file=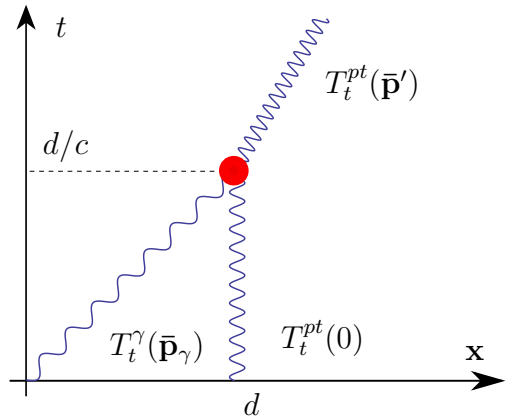,scale=0.93}
\caption{\small{
Causality for dynamically periodic fields. 
A massive field at rest in $| \mathbf{x}|=d$ with periodicity $ T^{pt}_{t}(0)$  ``absorbs''  after a  time delay $t = {d}/{c}$ the energy of  a massless field with periodicity $T^{\gamma}_{t}(\mathbf{\bar p}_{\gamma})$. 
After the interaction the resulting periodicity $T^{pt}_{t}(\mathbf{\bar p'})$ of the massive field is such that  $1/ T^{\gamma}_{t}(\mathbf{\bar p}_{\gamma}) + 1/  T^{pt}_{t}(0) = 1 / T^{pt}_{t}(\mathbf{\bar p'})$ due to energy conservation in the interaction point. 
}
\label{worldlinext4D}}
\end{center}
\end{wrapfigure}

 For instance, in fig.(\ref{worldlinext4D}) we suppose to turn on  a source with time periodicity $T^{\gamma}_{t}(\mathbf{\bar p}_{\gamma})$ in the origin of the axis,   
so that a massless field  
     is ``absorbed'' after a time delay by a massive field originally at rest. 
 The scenario is similar to the   Compton scattering 
   but the energy conservation among the quanta  can now be written as a conservation of the inverse of the periodicities between the fields involved in the interaction: $1/ T^{\gamma}_{t}(\mathbf{\bar p}_{\gamma}) + 1/  T^{pt}_{t}(0) = 1 / T^{pt}_{t}(\mathbf{\bar p'})$. 
  Through interaction,  the field passes from the original periodic regime to a different one.  
  Contrary to the static compactification scenario, this means that we can distinguish between an interaction before and an interaction after absorption. Thus we can give a time order to events and a dynamical compactification of the time dimension is compatible with relativistic causality.  
  The theory is therefore in agreement with special relativity and the notion of time is formally well defined. 
This remarkable result can be interpreted as a consequence of the fact that periodic (or antiperiodic) boundary conditions  
satisfy the variational principle, exactly as the usual boundary conditions with fixed values of the field, see eq.(\ref{bound:act:var}). 
As the Newton's law of inertial doesn't imply that non-isolated particles go on a straight line, our assumption of periodicity doesn't imply that a system of interacting elementary fields should appear to be periodic.

\section{Periodic mechanics}\label{Periodic:dynamics}

The previous results encourage an analysis of the mechanics of such  periodic fields which we expect to be non trivial due to self-interference. 
The theory   so far is analogous to finite temperature field theory - with  Minkowskian compact time - so it yields to a well defined field theory. 
For both massless and massive periodic fields we can explicitly write down the on-shell solutions of the equations of motion of the actions eq.(\ref{time:decompact}) or eq.(\ref{one:sum:act})  with the following notation \cite{Kapusta:1989tk}
 \begin{eqnarray}\label{hilbert:H}
  \Phi(\mathbf x,t) &=&  
    \sum_{n} \sum_{ \mathbf p_{n}}  a_n \phi_n(\mathbf x) u_{n}(t) 
  \end{eqnarray}
 where   
    $\sum_{\mathbf p_{n}}$  stands for the integral over  on-shell momenta and $a_n$ are the normalized coefficients of the Fourier expansion - see \cite{Dolce:2009ce} for more details.   
Being on-shell fields, after decompactification we find that the spatial components satisfy the  equations of motion
\begin{equation}
({\nabla}^2  +  k_n^2 ) \phi_{n}( x) = 0~. 
\end{equation} 
For the sake of simplicity in this section we suppose a single spatial dimension $x$. Both momentum and energy eigenmodes are orthogonal and complete. Using Poisson summation,\footnote{The Poisson summation implies that $\sum_{n}e^{-i n \alpha} = 2 \pi \sum_{n'}\delta(\alpha + 2 \pi n').$} the energy eigenmodes complete set is such that
\begin{eqnarray}\label{orthog:compl}
 \int_{0}^{\lambda_{x}}   \frac{d x}{\lambda_{x}}  \phi^{*}_{ n}( x) \phi_{m}( x) =  \delta^{}_{ n, m} ~~~, ~~~~~~~
 \sum_{n } \frac{ \phi^{*}_{n}({ x}) \phi_{n}( y)}{\lambda_{x}}=  \sum_{n'} \delta^{}({ x}-{ y} + \lambda_{x} n' )  ~.
\end{eqnarray}
The second relation shows that the spatial coordinate, similarly to the time coordinate, is defined modulo  $\lambda_{x}$ translations. 
Even though the whole information is in a single elementary space period $\lambda_{x}$,  we can always  write the above conditions extending the integration over the whole spatial region  $V_{x}$ where the field is supposed to be free and with an integer number of periods. 
In this case the  substitution is just $\int_{0}^{\lambda_{x}} {d x}/ \lambda_{x} \rightarrow \int_{V_{x}}^{} {d x}/ V_{x} $, where $V_{x} = N \lambda_{x}$ and $N \in \mathbb N$ is the integer number of periods in $V_x$. 

\subsection{Schr\"odinger equation and  Hilbert space }\label{Periodic:Paths}

The time evolution for the energy eigenmodes of the relativistic periodic field  eq.(\ref{hilbert:H}) is described by the ``bulk'' equations of motion 
\begin{equation}\label{Schrodinger:time}
(\partial_t^2 + \omega_n^2)u_{n}(t) = 0 ~,
\end{equation}
where the frequency spectrum is fixed by boundary conditions, eq.(\ref{peridic:BC}). It is given in eq.(\ref{massless:disp:relat}) and in eq.(\ref{massive:disp:relat}) for massless and massive periodic fields respectively.
These equations of motion along the time can be interpreted, together with the de Broglie relation eq.(\ref{fund:level}), as the  Schr\"odinger equation. 
Since the energy eigenmodes eq.(\ref{hilbert:H}) satisfy  
$ 
i \partial_t \phi_{n}( x) u_{n}(t) = \omega_n \phi_{n}( x) u_{n}(t)
$, 
we obtain indeed   the  Schr\"odinger equation for the field
\begin{equation}\label{Schrodinger:eq}
i \hbar \partial_t \phi_{n}({x}, t) = E_n \phi_{n}({x}, t) ~,
\end{equation}
which is the ``square root'' of the eq.(\ref{Schrodinger:time}), see also \cite{Dvali:2001gm,Carena:2002me}.
 Roughly speaking, this is due to the fact that the Klein-Gordon equation is indeed the relativistic generalization of the Schr\"odinger equation.

Another important point is that we are describing standing waves. Therefore this is the typical case where a  Hilbert space can be defined. Because of the orthogonality and completeness relations in eqs.(\ref{orthog:compl}) between the energy eigenmodes, it is  natural to introduce the following  inner product 
\begin{equation}\label{H:prod}
\left\langle \phi | \chi \right\rangle_{H} \equiv \int_{0}^{ \lambda_{x}} \frac{d x}{\lambda_{x}}  \phi^* ( x) \chi( x)~.
\end{equation}
This naturally yields to  an Hilbert space with the following eigenstates 
\begin{equation}\label{H:state}
\left\langle {x}| \phi_n  \right\rangle_{H} \equiv 
\frac{ \phi_n({x}) }{\sqrt{\lambda_{x}}}~.
\end{equation}
Furthermore, we can build the Hamiltonian operator as
\begin{equation}\label{hamilt:oper}
\hat{H} \left| \phi_n \right\rangle_{H}  \equiv \hbar \omega_n \left| \phi_n \right\rangle_{H} ~.  
\end{equation}
From the eq.(\ref{Schrodinger:time}) the time evolution for a generic state $\left|\phi(0)\right\rangle_{H}  = \sum_n  \alpha_{n}  \left|   \phi_n \right\rangle_{H} $ can now be written as  
\begin{equation}
\left|\phi(t)\right\rangle_{H}  = \sum_n e^{- i \omega_n t}\alpha_{n}  \left|   \phi_n \right\rangle_{H} ~, 
\end{equation}
that is, using the Hamiltonian operator \cite{Nielsen:2006th}, we can equivalently write 
\begin{equation}
\left| \phi(t)\right\rangle_{H}  = e^{-\frac{i}{\hbar} \hat{H} t}  \left| \phi (0) \right\rangle_{H}   ~.
\end{equation}
The Schr\"odinger equation can be written in a more familiar form
 \begin{equation}\label{Schrodinger:equation:formal}
 i \hbar \partial_t\left| \phi(t)\right\rangle_{H} = \hat{H} \left| \phi(t)\right\rangle_{H} ~.
 \end{equation} 
 
We are assuming that the operator $\hat{H}$ is not a function of time  
 (no source terms and no interactions in order to preserve  periodicity inside the volume $V_x$). It corresponds formally to the generator of  time translations
\begin{equation}\label{evol:oper}
U(t',t) = e^{-\frac{i}{\hbar} \hat{H}(t-t')}~.
\end{equation}
This time evolution between generic $t'$ and $t''$ can be justified by  complex dynamics caused by the periodic time dimension 
and it has the Markovian operator property 
\begin{equation}
U(t'',t)U(t,t')= U(t'',t')~; ~~~~~~~t'' \geq t \geq t'~.
\end{equation}
 Using this property we divide the time interval in $N$ elementary intervals of  length $\eps$ 
\begin{equation}\label{markov}
U(t'',t') = \prod_{m=0}^{N-1} U(t'+  t_{m+1}, t' + t_{m} -  \epsilon)~;~~~~~~~N \eps = t'' - t'~,
\end{equation} 
where we are using the notation $t_{m+1}= (m + 1)\eps$ and  $t_{m}= m \eps$. 
Notice that for an Euclidean time, $U(t + \epsilon, t)$ is  analogous to the transfer matrix of classic statistical mechanics, so a statistical interpretation of the periodic dynamics is possible. 

\subsection{Feynman path integral}\label{Paths:integral}

We point out that, without any further assumption than periodicity, all the ingredients to build a path integral are already contained in this periodic field theory: we have a Hamiltonian time evolution operator eq.(\ref{evol:oper}), with the Markovian property eq.(\ref{markov}) and a complete set of energy eigenfunctions  eqs.(\ref{orthog:compl}).  
From a mathematical point of view and proceeding  completely in standard way, we make use of the completeness and orthogonality relations of the $\phi_{n}({x})$  in  eq.(\ref{markov}).   Separating the space-time  evolution in infinitesimal parts we get
\begin{eqnarray}\label{int:evol:oper}
U({x}'', t''; {x}', t' ) &=& \!\! \int_{0}^{ \lambda_{x}} \!\! \left( \prod_{m=1}^{N-1}  \frac{d x_m}{ \lambda_{x}}  \right )
U({x}'', t"; {x}_{N-1},   t_{N-1}  ) U({x}_{N-1}, t_{N-1}; {x}_{N-2},   t_{N-2} )  \nn \\
&\times& \dots \times  
  U({x}_{2}, t_{2}; {x}_{1}, t_{1} ) U({x}_1, t_{1} ; {x}', t'  )~.
\end{eqnarray}
The  elementary periodic evolutions  between  spatial points ${x}_k = {x}(t_k)$ to ${x}_{k-1} = {x}(t_{k-1})$ turn out to be given by
\begin{eqnarray}\label{microevolut}
U({x}_{m+1}, t_{m+1}; {x}_{m}, t_{m} ) &=& \sum_{n_{m}}  e^{- \frac{i}{\hbar} ( E_{n_{m}} \Delta \eps_{m} -  {p}_{n_{m}}  \Delta {x}_m)}  ~, 
\end{eqnarray}
with the notation $\Delta x_{m} = x_{m+1} - x_{m}$ and $\Delta \eps_{m} = t_{m+1} - t_{m}$. As already mentioned  the energy spectrum is $E_{n}({\bar{p}}) = n \hbar \bar \omega({\bar{p}})$. 
These elementary space-time evolutions  correspond to the ``unitarized" periodic field $ \phi(\Delta x_{m}, \Delta \eps_{m})$  (that is to say a periodic field with  unitarized coefficients $ a_{n} \equiv 1$, $\forall n$). Using Dirac notation eq.(\ref{H:prod}) - see also the definition of the expectation value eq.(\ref{mean:value}) in the next section - we get the familiar form
\begin{eqnarray}
U({x}_{m+1}, t_{m+1}; {x}_{m}, t_{m} ) =  \phi(\Delta x_{m}, \Delta \eps_{m}) &=&  \lambda_{x} \left\langle \phi  \right| e^{-\frac{i}{\hbar}( \hat{H} \Delta \eps_{m}  - \hat p \Delta x_{m} ) }  \left| \phi \right\rangle_{H}~,
\end{eqnarray}   
where the operator $\hat p$ is defined in analogy with  the Hamiltonian operator in eq.(\ref{hamilt:oper}).
Plugging these microscopic evolutions in eq.(\ref{int:evol:oper}) we get formally  the Feynman  path integral in phase space for a time independent Hamiltonian 
\begin{equation}\label{periodic:path.integr:Oper:Fey}
 U({x}'', t''; {x}', t' )   = \lim_{N \rightarrow \infty}   \int_{0}^{\lambda_{x}} \left ( \prod_{m=1}^{N-1} {d x_m}{ } \right ) \left \lbrace \prod_{m=0}^{N-1} \left [   \left\langle \phi  \right| e^{-\frac{i}{\hbar}( \hat{H} \Delta \eps_{m}  - \hat p \Delta x_{m} ) }  \left| \phi \right\rangle_{H} \right ] \right  \rbrace ~.
\end{equation}
Remarkably this fundamental result  has been obtained just assuming relativistic periodic waves without any further assumption such as commutation relations between the canonical variables. 
 We will see in sec.(\ref{Commutation:relations}) that commutation relations  can be derived directly from periodic fields, but it can be obtained from this path integral as well. 

Assuming that in the non-relativistic limit only the first mode ($n=1$) is largely populated it is possible to derive the non-relativistic free particle limit of the theory, see par.(\ref{Massive:bosons}) and  \cite{Dolce:2009ce} for more details.

 The path integral formulation  arises as a direct consequence of  the fact that  
 the non trivial periodic dynamics yield  a class of equivalence between  initial  and final  points translated by space-time periods. 
  It is possible to reach a given final configuration following a class of periodic on-shell paths, \textit{i.e.} paths with different winding numbers. 
     In other words, contrarily to the Feynman formulation where there is a unique classical path, a periodic field  interferes with itself because of the periodic boundary conditions and the consequent equivalence class of paths with different winding numbers, without relaxing the validity of the least action principle  \cite{Dolce:2009ce}. 

\subsection{Commutation relations}\label{Commutation:relations}

 In sec.(\ref{Periodic:Paths}) we  found that periodic fields can be written in a Hilbert space with   time evolution given by the Schr\"odinger equation. Now we would like to have commutation relations  in order to formalize the analogy with the canonical  formulation of quantum mechanics  as well. 
Looking at the inner product in eq.(\ref{H:prod}) we identify the mean  value of a given observable $F(x)$ between generic initial and final states  $|\phi \rangle $ and $|\chi \rangle $ as 
\begin{eqnarray}\label{mean:value}
& & \langle \chi(x_{f}, t_{f})| F(x) | \phi(x_{i},t_{i}) \rangle_{H} = \nn \\
&=& \int_{0}^{\lambda_{x}} \frac{d^{} x}{\lambda_{x}} \sum_{n, m}  \alpha^{*}_{\chi_m} e^{ i \omega_{\chi_m} t_{f}   - i  k_{\chi_m} (x_{f} - x)}  F(x)  e ^{- i \omega_{\phi_n}  t_{i}  + i  k_{\phi_n} (x - x_{i})}   \alpha_{\phi_n}   ~,
\end{eqnarray} 
where $\lambda_x$ is the spacial period in $x$ - the integration volume can be extended to the whole  periodic region $V_{x}$. 
To determine commutation relations we follow \cite{Feynman:1942us}, but using directly the unitarized periodic fields $\phi(x,t)$ rather than the periodic path integral.
In fact \cite{Dolce:2009ce}  there is an equivalence between the two formulations. 
We continue by evaluating the mean value of $\ptl_{x} F(x)$. 
Integrating by parts eq.(\ref{mean:value}) and considering 
the periodicity $\lambda_x$ of the spatial variable and of the states,\footnote{It is equivalently possible to assume Dirichlet boundary conditions at the boundaries $\Phi(0,t)=\Phi(\lambda_x,t)\equiv 0$ instead of periodicity. In this way the only inessential difference is that there is no zero mode $n=0$, that is no translational mode.}  we get 
\begin{eqnarray}
\!\!\!\! & &\langle \chi(x_{f}, t_{f})| \ptl_{x} F(x) | \phi(x_{i},t_{i}) \rangle_{H} =   \nn \\
 &=&  \frac{i}{\hbar}\!\int_{0}^{\lambda_{x}}\!\! \frac{d^{} x}{\lambda_{x}} \sum_{n, m}  \alpha^{*}_{\chi_m}   \alpha_{\phi_n} e ^{ i \omega_{\chi_m} t_{f}  -  i  k_{\chi_m}\! (x_{f} - x)}  [p_{\chi_m }  F(x)\! - \! F(x) p_{\phi_n}   ]  e ^{- i \omega_{\phi_n}   t_{i}  + i  k_{\phi_n} \!(x - x_{i})}    \nn \\
 &= &\frac{i}{\hbar} \langle \chi(x_{f}, t_{f})| \hat p  F(x) -  F(x) \hat p    | \phi(x_{i},t_{i}) \rangle_{H} ~.
\end{eqnarray}
 Thus, by choosing  $F(x) \equiv x$, it turns out that
\begin{eqnarray}
\langle \chi(x_{f}, t_{f})| 1 | \phi(x_{i},t_{i}) \rangle_{H}   
 = \frac{i}{\hbar} \langle \chi(x_{f}, t_{f})| \hat p  x -  x \hat p    | \phi(x_{i},t_{i}) \rangle_{H} ~,
\end{eqnarray}
which, for generic initial and final states, reproduces 
the commutation relation of quantum mechanics 
\begin{equation}\label{comm:rel}
[x,\hat p] = i \hbar	~.
\end{equation}
We thus find  that, besides the Feynman formalism derived in the previous subsection,  the fundamental elements to build the canonical formulation of quantum mechanics  are already  contained in this theory as well.\footnote{More easily we note that $[x, -i \hbar \partial_x]\Phi(x,t)= i \hbar \Phi(x,t)$. }

\subsection{Heisenberg uncertainty relation}\label{Heisenberg:uncertainty:relation}

 For a periodic wave it is possible to obtain an  uncertainty rule in a rather immediate and trivial way. 
 To determine the frequency of a free wave  and thus the energy of the related mode  we must count the oscillations for at least a  time interval greater than the fundamental period: the longer the measuring time, the lower the frequency uncertainty.
  Mathematically we can see this by noting that the  phase  $ \bar E t / \hbar $ is defined modulo factors $2 \pi n$. 
  Supposing for simplicity $n = 1$, we can reabsorb this factor either  as a variation of the time variable $\Delta t = 2 \pi \hbar / \bar E$ or  of the energy $\Delta E = 2 \pi \hbar / t$, so that 
$
  \Delta E \times \Delta t = {(2 \pi \hbar)^{2}}/{ \bar E t}
 $, 
   which is minimized by the largest value of the time in the denominator $t \rightarrow T_{t}$. 
  Finally, we recover the Heisenberg uncertainty relation\footnote{Taking into account the square modulo of the wave function we have a  phase invariance  $ n \pi$ which gives a  factor $1/2$ in the final result.}
 \begin{equation}\label{indeterm}
 \Delta E \times \Delta t  \geq 2 \pi \hbar = h~.
 \end{equation} 
 This is a direct  consequence  of the de Broglie assumption in eq.(\ref{fund:level}), that can be generalized  \cite{Nielsen:2006th} to
\begin{equation}\label{bohr-sommerfeld}
E_n {R_{t}} = n\hbar~.
\end{equation}
This relation can be regarded as the semi-classical Einstein's formulation of the Bohr-Sommerfeld quantization condition:  in a given potential only  phase-space orbits which fit in an integer number of periods $T_{t}$ are allowed. This simple recipe is sufficient to solve many problems of non-relativistic quantum mechanics\footnote{Modulo the zero-point energy which must be fine-tuned using twisted boundary conditions.}, such as the quantum harmonic oscillator, the anharmonic or anisotropic quantum oscillator,  linear potential, of the various well potentials and Dirac delta potentials, the hydrogen atom, etc...    \cite{Dolce:2009ce}.

\section{Quantum mechanical interpretation}\label{QM:interpret}

Since we have inferred  the Hilbert space eqs.(\ref{orthog:compl}), eq.(\ref{H:prod}) and eq.(\ref{H:state}),  the Schr\"odinger equation eq.(\ref{Schrodinger:eq}), eq.(\ref{hamilt:oper}) and eq.(\ref{Schrodinger:equation:formal}),   the commutation relations eq.(\ref{comm:rel}), the  path integral eq.(\ref{periodic:path.integr:Oper:Fey}),   the  Heisenberg uncertainty relation eq.(\ref{indeterm}) and the Bohr-Sommer\-feld condition eq.(\ref{bohr-sommerfeld}) from the periodicity assumption, it is reasonably correct to interpret our theory as a quantum theory.

In general, the standard fields can be thought of as an integral over elementary harmonic oscillators with angular frequencies $\bar \omega(\mathbf {\bar p})$. 
In the usual formulation a non interacting classical field  with fixed momentum $\mathbf{ \bar{p}}$ is a single de Broglie plane wave with fixed frequency $\bar \omega(\mathbf {\bar p})$. Therefore it can be described in terms of a single harmonic oscillator with characteristic periodicity $T_{t}(\mathbf {\bar p}) = 1 /\bar \omega(\mathbf {\bar p})$. Its angular frequency $\bar \omega(\mathbf {\bar p})$  must be written as in eq.(\ref{massless:disp:relat}) or eq.(\ref{massive:disp:relat}) respectively for massless or massive fields.   

The usual quantization of  bosonic fields (namely the  second quantization) 
is obtained by explicitly quantizing each harmonic oscillator, that is by imposing commutation relations. 
After  normal ordering, every single harmonic oscillator  has a quantized energy spectrum
$ 
\text{:}E_{n}(\mathbf {\bar p})\text{:}  =  \text{:}\hbar \omega_{n}(\mathbf {\bar p})\text{:}  =   n \hbar  \bar \omega(\mathbf {\bar p})
$.\footnote{According to the Born rule, we assume that  the probability density $\rho = |\Phi(\mathbf x, t)|^2 $ associated to the  periodic fields   
is given by the inner product  $\langle \Phi(\mathbf x, t)| \Phi(\mathbf x, t) \rangle_H$, eq.(\ref{H:prod}). It is interesting to note that $\rho$ corresponds to the non-relativistic limit of the charge density $j_0$ related to the periodic field $\Phi(\mathbf x, t)$ ({for instance we may also note that when we observe a particle we inevitably stop it on the rest frame of the detector} \cite{Rosenstein:1985wb,Nikolic:2007id,Nikolic:2008sn}) - see for more details \cite{Dolce:2009ce}.  } 
These are just the admitted energies of a periodic field with  periodicity $T_{t}$  as prescribed by eq.(\ref{fund:level}). 
On the other hand, all the arguments given so far can easily be generalized to the orbifold case  $t \in \mathbb{S}^1/\mathcal{Z}_2$, which  gives the spectrum with vacuum energy 
$ 
E_{n}(\mathbf {\bar p}) =  \left(n + {1}/{2}\right) \hbar \bar \omega(\mathbf {\bar p})
$ 
by supposing that the  field is odd under the $\mathcal{Z}_2$ parity, that is antiperiodicity (because of the analogy with finite temperature field theory and for the scope of this paper we can associate this odd orbifold to fermionic fields in order to satisfy the spin\-statistics relation \cite{Kapusta:1989tk}). 
Because of the similarities with finite temperature field field theory, for the scope of this paper  fermions can be thought of as antisymmetric fields.   
Similarly,  a generic value of the vacuum energy $v \hbar \bar \omega$ can be obtained by assuming twisted periodic boundary conditions $\Phi(\mathbf{x},t)= \exp{[-2 \pi v]} \Phi(\mathbf{x}, t + 2 \pi R_t)$.
Anyway, as explained in more detail in the sec.(\ref{determinism})  and  \cite{'tHooft:2006sy,'tHooft:2007xi,Dolce:2009ce}, these contributions to the energy 
are of "little importance" since they come  from  phase factors in front of the fields.  
Furthermore, we point out that  "the Casimir effect, often invoked as decisive evidence that the zero point energies of the quantum field are real, [...] can be formulated and the Casimir forces can be computed without reference to zero point energies" \cite{Jaffe:2005vp}. Indeed they can be formulated in a classical way in terms  Van der Waals  forces \cite{Jaffe:2005vp}  between the electrons in the two metallic plates or using boundary conditions on the metallic plates. 
 Because of the analogy with finite temperature field theory and for the scope of this paper we can associate this odd orbifold to fermionic fields in order to satisfy the spin-statistics relation \cite{Kapusta:1989tk}. 
Indeed, under this hypothesis fermionic fields have  vacuum energy $ \hbar \bar \omega(\mathbf {\bar p})/2$.

We  summarize the analogy between periodic fields and quantum fields that we want to explore by saying that every relativistic field $\Phi(\mathbf x, t)$ with assigned four-momentum $ {\bar p}_\mu$ has a fixed space-time periodicities  $T_{\mu} = h / \bar p^\mu$, in agreement with the de Broglie hypothesis.  
It  can be decomposed in  a series of eigenstates  $\phi_{n}(\mathbf p)$ with energies $n \hbar \bar \omega(\mathbf {\bar p}) $ whose interpretation is  in terms of the    
 ``quanta''  of the related quantum field. 
 Further evidencies for this mapping with ordinary quantum mechanics are given in  \cite{Dolce:2009ce} where we describe the essential phenomenology and elucidative applications.

\subsection{Determinism}\label{determinism}

Another important aspect which motivated the investigation upon periodic time dimension is the 't Hooft determinism: there is a ``close relationship between the quantum harmonic oscillator and a classical particle moving along a circle" \cite{'tHooft:2001fb,Elze:2002eg,'tHooft:2001ar}. 
We approach the 't Hooft determinism by assuming periodic  fields with time period $T_{t}$ on a lattice with $N$ sites, in order to (de)construct \cite{Arkani-Hamed:2001ca,Berezhiani:2002et} the time dimension.\footnote{In \cite{Berezhiani:2002et}   a four-dimensional Yang-Mills field theory  emerges dynamically by  dimensional (de)construction mechanism \cite{Arkani-Hamed:2001ca,Berezhiani:2002et} applied to replicated three-dimensional gauge theories (a moose model in three-dimensions). 
This  dynamically constructed periodic time dimension leads to  the Heisenberg uncertainty relation eq.(\ref{indeterm}) and energy quantization eq.(\ref{discr:ener:lev}).}
 We  associate to every discretized phase, \textit{i.e} to every site of the lattice, a column state ${|0 \rangle, |1 \rangle, \dots, |N-1 \rangle}$. 
 The model is analogous to an harmonic system of $N$ masses and  springs on a ring. 
  It turns out that if the time accuracy is $\Delta t \gg T_{t}$, at every observation the field $\Phi(x, t)$ appears in an arbitrary phase $|n \rangle$ of its cyclic evolution, so that the evolution has  an apparent aleatoric behavior; as if observing a clock under a stroboscopic light \cite{Elze:2002eg}, or a dice rolling to fast to predict the result.  
In fact,  as already discussed in sec.(\ref{Massive:bosons}), if the underlying periodic dynamics are  too fast to be observed ($\lesssim 10^{-20} s$),
the time evolution between two column states $|n \rangle$ can only be described statistically through the operator $\tilde U(\Delta t = \eps) = \exp[{-\frac{ i }{\hbar} \tilde H \eps}]$, where $\tilde H$ is a $N \times N$ matrix. 
In the limit of large $N$ the column states obey to the relation 
$ 
\tilde H |n \rangle \sim \hbar \bar \omega \left(n + {1}/{2}\right) |n \rangle 
$. 
This  reproduces just the energy eigenvalues of the quantum harmonic oscillator, apart for a phase factor of ``little importance'' in front of the operator $\tilde U(\eps)$  which reproduces the factor $1/2$ in the eigenvalues \cite{'tHooft:2001ar,'tHooft:2001ct,'tHooft:2006sy,'tHooft:2007xi}.  
From the evolution operator $\tilde U(t)$  we can once again  observe the analogy between quantum and statistical mechanics.\footnote{Here we note  also a resemblance with the random walk problem which was originally solved using its analogy with interference of iso-periodic waves with random phase distribution \cite{Rayleigh:1905}.}
Due to the extremely fast underlying dynamics  we loose information about the fundamental classical theory which give rise to the quantum behavior. 
For this reason we can speak about deterministic or pre-quantum theories.\footnote{Further similitudes  with the 't Hooft determinism are given by the fact that the de Broglie time periodicities can be regarded as ``cellular automata" \cite{hooft-2009-0} and our fields, being constrained in a periodic time dimension, share interesting analogies with black-hole thermodynamics \cite{hooft-2009-1}.}  
 It is interesting to point out that, since  a  periodic time dimension can induce  periodicity to the proper time, that is the worldline parameter of the fields,  we have an analogy to  string theory where one of the two worldsheet parameters is compactified. 

 Motivated  by the 't Hooft determinism and the attempts to quantize gravity, a model of a classical particle moving in five-dimensions, two of which are compactified on a torus,  is illustrated  in \cite{Elze:2002eg,Elze:2003nu,Elze:2003ws,Elze:2003tb,Elze:2005gv}.  
 The ergodic dynamics associated to this model  give rise to an effective time and thus to a so called ``stroboscopic quantization''. 
 The relevant idea here is  the derivation of a notion of ``time'' which emerges from  the ``ticks'' of an ergodic system. 
 Similarly, in our theory the notion of time emerges from the ``ticks'' provided by the de Broglie internal clocks.  
 Geometric quantization \cite{Gozzi:1991di,Gozzi:2006cv} seems to indicate another connection between the notion of time and  quantization. 
 In fact, in this theory some quantum phenomena emerge by integrating out two grassmannian partners  of the physical time.

In the 't Hooft approach to determinism as well as in the model with ``stroboscopic quantization'' there is the attempt to avoid local hidden variables.  
It is worth noting that the approach with compact time  proposed throughout this paper has  not  local-hidden-variables  that must be integrated out to get the quantum observables. We have just space and time coordinates which are physical variables. On the other hand the periodic conditions in eq.(\ref{peridic:BC}) can be regarded as an element of non locality (which is consistent with relativistic causality) in the theory. 
Therefore  model proposed  in this paper is deterministic\footnote{Here we mean mathematical determinism. From a practical point of view is in fact impossible to measure time with an infinite accuracy and thus to know the exact boundary conditions of the system under investigation, see sec.(\ref{Heisenberg:uncertainty:relation}).} since it represents a possible way out of the  Bell's inequality or  similar non-local-hidden-variable theorems \cite{Nikolic:2006az}.

\subsection{Compact space-time formalism}

Paraphrasing the Newton's law of inertia and the de Broglie hypothesis of periodicity, we assume that every isolated elementary system (every free elementary field) has persistent and constant time periodicity (as long as it doesn't interact) fixed by the inverse of the energy  $T_t = h / \bar E$. Considering the periodicities induced on the modulo of the spatial dimensions, the resulting space-time periodicities are  those of the ordinary de Broglie waves and therefore they are consistent with special relativity.   

A conceptual effort is required for a conceptual understanding of this theory, because it adds a property of periodicity  to our ordinary notion of relativistic time. 
 From a formal point of view,  in this relativistic theory the physical time is well defined  through the relation between periodicity and energy. 
 It respects all the  required properties such as Lorentz transformations, causality and chronological ordering. 
But, as much as the Newton's law of inertia doesn't imply that every point particle goes in a straight line, our assumption of periodicity does not  mean that the physical world should appear to be periodic. 
In fact there is not a single static periodicity which would serve as privileged reference. 
On the contrary elementary systems (that we  represent as fields) at different energies have different periodicities.\footnote{This concept has a precise mathematical justification, in fact Fourier showed that every regular  (not necessarily periodic) function  can be expressed as an integral over periodic functions.} 
The conjecture is that the combination of these different periodicities, that for massless fields  may effectively vary between  the Planck time $\sim 10^{-44} s$  to  the age of the Universe $\sim   10^{15} s$ or more (in the hypothesis of a cyclic universe), is the reason of our perception  of the time flow. 
Furthermore, through interactions the elementary systems pass from a periodic regime to another periodic regime, forming in general ergodic and even more chaotic evolutions. This give rise to a possible statistical interpretation of the arrow of time. 
 
To figure out the possibility of a formulation of relativistic fields in compact space-time dimensions we follow few simple logical steps. 
 Ordinary field theory is based upon de Broglie waves that are then quantized by imposing commutation relations. 
To every de Broglie wave there is associated a frequency proportional to its energy and thus an intrinsic periodicity which is usually called de Broglie internal clock. 
In fact time can be only defined by assuming periodicity, in order to ensure that the duration of a unit of time is always the same; in past, in the present and in the future.  
Our usual - non compact - time axis is defined with reference to the Cs-133 atomic clock whose period is about $10^{-10} s$, an electron at rest has an internal de Broglie clock of about $\sim 10^{-20} s$ whereas an  hypothetical heavy particle of $1$ TeV has an internal clock of $\sim 10^{-27} s$. 
Depending on its energy, a massless field such as the electromagnetic field (or the gravitational field),  can in principle have all the possible intrinsic periodicities. In particular it can have an infinite period (or of the order of the age of the universe). 

Every value of our time axis is characterized by a unique combination of phases of all the de Broglie clocks of the elementary fields constituting the system under investigation. 
This means that the external time axis can be  dropped and the flow of time can be effectively described using the de Broglie internal clocks as in a calendar or in an stopwatch - the massless fields provide the long time scales. 
This is a simplified picture since we must remember that the clocks can vary periodicity through interaction (exchange of energy), that  periods depend  on the reference systems according to the relativistic laws and that the combination of two clocks with irrational ratio of periodicities gives ergodic (not exactly periodic) evolutions. 
It is interesting to note that this picture is of particular interest for the problem of the time symmetry in physics, in fact the de Broglie clocks can be equivalently supposed to be clockwise or anticlockwise. 
Remembering the Einstein's definition of relativistic clock \cite{Einstein:1910} (see introduction), we can restrict our attention on a single period of every de Broglie internal clock, that is of every elementary field constituting our system. 
This means the physical information of the fields is contained in the single periods, therefore we formalize this by investigating fields with compact time and periodic (or Dirichlet) boundary conditions. 
Similar argumentations hold for the spatial dimensions.  In the non relativistic limit, matter fields can be approximated as with infinite spatial periodicity and microscopic time compactification proportional to its Compton wavelengths.  
Hence they can be regarded as nearly three spatial dimensional objects. Furthermore, since they are spatially  localized inside their  microscopical Compton wavelengths, they can be effectively  regarded as non-relativistic point-like particles.    

Another intuitive image can be found in  the many similarities with acoustic waves \cite{johnson:227}. 
  The sound is a set of standing waves generated by a string, a membrane or a solid body vibrating in one, two or three compact spatial dimensions respectively. 
  The harmonics (frequency eigenstates) of these acoustic waves are those allowed by the size of the spatial compact dimensions in which the sound source is embedded. In a full relativistic generalization of the sound waves, our relativistic fields can be thought of as being generated by vibrating objects (sources) characterized by intrinsically compact space-time dimensions. Roughly speaking, massless fields at small momentum have nearly infinite time periodicity (nearly continuous energy spectrum) so that they act like sound fields in a medium whereas matter fields, even at small momentum, have compact time dimension and they act like sound sources. The difference with the usual  field picture is that now we allow a ``timbre" to the de Broglie waves, that is we consider all the frequencies, and thus the different spectral compositions, allowed by the space-time periodicities $2 \pi R_\mu$. 

\subsection{Towards a formalization of interactions}
  
So far we have illustrated the formal and conceptual correspondences between a field theory with periodic time dimension and the usual quantum theory, concerning only with free field. 
The exact solution of  the interaction between periodic  fields and thus the transition between to different periodic regimes  is beyond the scope of this article. 
 Most likely, it would require the development of  a perturbative theory, adding an interaction term to the Hamiltonian of  the periodic path integral eq.(\ref{periodic:path.integr:Oper:Fey}). 
  
 To give  a qualitative picture of the interacting periodic fields, 
    the most trivial example is  Compton scattering $e' + \gamma' \rightarrow e'' + \gamma''$. 
  As already mentioned about fig.(\ref{worldlinext4D}), we must merely consider  the energy-momentum  conservation  in terms of conservation of the inverse of the space-time periodicities of the fields involved, $1/ T_{\mu}^{\gamma'} + 1/ T_{\mu}^{e'} = 1/  T_{\mu}^{\gamma''} + 1/  T_{\mu}^{e''}$. 
  The  change in periodicity of a field during the interaction (that is when the field has significative overlaps or interference) can be regarded as a deformation of the space-time compactification lengths.
  This problem can be equivalently reformulated by imposing a deformation of the metric. Hence interactions can be interpreted in terms of relativistic geometrodynamics since this argumentation leads to  field theories on curved space-time. 
 For instance \cite{Dolce:2009ce}, as we will expose in a dedicated paper,  we can imagine to prepare a volume of quarks and gluons at  high energy, for instance using a collider.  
 The system looses energy by radiating hadronically or electromagnetically  \cite{Satz:2008kb}.  
 In first approximation,  
 as predicted by the hydrodynamic Bjorken model \cite{Magas:2003yp} and in similitude with thermodynamic system \cite{Satz:2008kb}, 
 the quark-gluon plasma  passes exponentially from a high energy regime characterized  by  small periodicities, to a low energy regime characterized by  large periodicities. 
 This conformal exponential dilatation of the space-time periodicities  turns out to be described in terms of five-dimensional fields with zero five-dimensional masses embedded in a ``virtual" AdS metric, similarly to sec.(\ref{Massive:bosons}).
From the mapping with quantum mechanics described so far, by imposing such a dilatation of the periodicities we expect to observe an evolution of the quantum observables with the energy. 
Indeed it turns out that the gauge coupling has a logarithmic running with the energy \cite{Pomarol:2000hp,ArkaniHamed:2000ds}. 
In fact, it is well know that the classical correlator of a classical field in a warped background can be approximatively matched with the quantum two point function of QCD. 
  Hence, we get a close parallelism with the AdS/CFT correspondence which originally motivated our study. 
  I fact, interpreting  the Maldacena conjecture \cite{Maldacena:1997re}  as in  Witten's work \cite{Witten:1998qj}, it describes  a parallelism between classical fields  in a warped dimension and  quantum  phenomena in a lower dimensional conformal theory, that is it encodes the quantum behaviors  in  classical configurations of fields in an (warped) extra-dimension.

\addcontentsline{toc}{section}{Conclusions} 
\section*{Conclusions}

We  investigated the hypothesis of  dynamical and local space-time periodicities, extending the ``old  quantum theory".  
Since these periodicities are the natural de Broglie periodicities of the classical fields, the resulting theory respects Lorentz invariance, preserves causality, allows time ordering, and  reproduces the relativistic field theory  in the limit of no boundaries. 
In fact,  periodic conditions imposed to the relativistic waves, similarly to the usual boundary conditions, minimize the relativistic action.
Indeed, special relativity prescribes that  time is  a local and dynamical property.
In  fields with compact space-time dimensions  this property is manifest through the inverse proportionality between the energy and the time periodicity. 
We found that a massive periodic field, whose characteristic rest spatial width is its  Compton length, has a extremely fast intrinsic periodicity ($\lesssim 10^{-20} s$) fixed by the inverse of its mass, as conjectured by de Broglie.  
 The space-time periodicities are different if observed from different inertial frames, in agreement with Lorentz transformations and relativistic dispersion relations. Hence, the theory is covariant.

The study of the compactification of the time dimension  has highlighted  remarkable  connections between  relativistic,  quantum  and thermal theories.  
We  pointed out several remarkable correspondences to the usual quantum theory such as the arising of a  discretized energy spectrum, of commutation relations and of  uncertainty  relations. 
 The effective time evolution of a periodic field is described by the Schr\"odinger equation in a Hilbert space. 
  Due to invariance by space-time translations of periods there are  different classical trajectories with different winding number between the initial and final points. This gives rise to  interference between different on-shell paths and thus to a path integral formulation, without relaxing the variational principle. 
  As a consequence of the periodic nature of the fields, typical quantum phenomena such as  black body radiation, the double slit experiment, Schr\"odinger problems, superconductivity, and many others can easily be reformulated.  
 The connection with thermal theory comes because of the close analogy with the finite temperature field theory, and because of the underlying statistical laws.
   Indeed we have tried to construct a consistent description of these three theories using the simplest physical system possible, essentially waves with boundary conditions.

 
 The field theory proposed here is a good candidate for pre-quantization since quantization arises from a deterministic theory instead of being imposed. 
  As the AdS/CFT correspondence, which seems to have an immediate interpretation in this theory, the results obtained so far  are  non trivial. 
They seem to open a new scenario where  a  compact time dimension arises as something more physical than a  simple mathematical trick, as believed in finite temperature field theory.
Indeed, a dynamically compact  Minkowskian time leads to the  concrete possibility to combine special relativistic and quantum theory  in a  deterministic wave theory. 
 The  great advantages of such deterministic theory can be potentially extended to all the quantum mechanical applications but especially in those  branches of physics where the quantum and relativistic mechanics  are difficult to conciliate, such as some aspects of high energy quantum field theory  and  quantum gravity.

The concept of time arising from this theory satisfies all the requirements prescribed by special relativity and,  combining the different de Broglie internal clocks of the elementary fields as in a calendar or in a clock, we can indeed fix and order events in time. This approach is of particular interest for the problem of the time symmetry in physics.    
The non periodic phenomena that we observe can be easily explained by the fact that systems can pass from a periodic regime to another through interactions  (energy exchange). 
If non periodic systems or similarly systems with periodicities larger than our observation time are  interacting with the elementary system  we are measuring, its  periodic evolution will be no more  manifest.\footnote{For instance the universe can be cyclic or not,  and with respect to this master time scale more and more events appear to have or have not a periodic nature.} 

Time has been defined by counting the number of oscillations of the Cesium atom or  of the incense lamp of the Pisa Dome,  the number of the orbits of the Earth or of the Moon. 
But all these definitions inevitably make use of the \textit{a priori} assumption of periodicity of isolated elementary systems and, ``by the principle of sufficient reason",  we  assume that the whole information of these elementary systems  is encoded in a single period, as implicitly said by Einstein himself \cite{Einstein:1910} in his definition of a relativistic clock. 
For this reason, and  for the ones mentioned in this work, we consider it worth investigating the physical consequences of an intrinsically cyclic nature of time.

\addcontentsline{toc}{section}{Acknowledgements}

 \section*{Acknowledgements} 
 I would like to thank in chronological  order D. Dominici, K. Schilcher, V. Ahrens,  M. Neubert, M. Reuter, N. Papadopoulos, F. Scheck and E. Manrique   
 for the fruitful and fair discussions on this ticklish topic and their encouragements. I am  grateful to V. Ahrens, F. Scheck, T. Pfoh, E. Manrique, B. Pecjak, L.L. Yang, and M. Bauer H.  for the critical reading of the manuscript and all the THEPer for listening patiently the long presentations of this work. The help and the support of V. Ahrens has been essential, not only in writing.  
 
 This work  was started in the in the Department of Physics and INFN of the Florence University, Italy, during my PhD  and in part financially supported by the Fondazione Angelo  della Riccia  with the hospitality of the IFAE in the Universitat Aut\`onoma de Barcelona, Spain.

\bibliographystyle{utphys}
\bibliography{comp3+1}

\providecommand{\href}[2]{#2}\begingroup\raggedright\begin{thebibliography}{10}

\bibitem{Einstein:1910}
A.~Einstein, ``Principe de relativit\'e,'' {\em Arch. Sci. Phys. Natur.} {\bf
  29} (1910).

\bibitem{Zeh:1992vf}
H.~D. Zeh, ``{The Physical basis of the direction of time},''. Berlin, Germany:
  Springer (1992) 188 p.

\bibitem{Zee:2003mt}
A.~Zee, ``{Quantum field theory in a nutshell},''. Princeton, UK: Princeton
  Univ. Pr. (2003) 518 p.

\bibitem{Zinn-Justin:2000dr}
J.~Zinn-Justin, ``Quantum field theory at finite temperature: An
  introduction,''
\href{http://www.arXiv.org/abs/hep-ph/0005272}{{\tt hep-ph/0005272}}.

\bibitem{Kapusta:1989tk}
J.~I. Kapusta, ``Finite temperature field theory,''.

\bibitem{Penrose:2008}
R.~Penrose, ``{Conformal boundaries, Quantum Geometry, and Cyclic
  Cosmology},''. Prepared for Beyond Einstein 2008 Conference, Mainz, Germany,
  22 - 26 Sept 2008.

\bibitem{Steinhardt:2002kw}
P.~J. Steinhardt and N.~Turok, ``{The cyclic universe: An informal
  introduction},'' {\em Nucl. Phys. Proc. Suppl.} {\bf 124} (2003) 38--49,
\href{http://www.arXiv.org/abs/astro-ph/0204479}{{\tt astro-ph/0204479}}.

\bibitem{Nielsen:2006vc}
H.~B. Nielsen and M.~Ninomiya, ``{Intrinsic periodicity of time and non-maximal
  entropy of universe},'' {\em Int. J. Mod. Phys.} {\bf A21} (2006) 5151--5162,
\href{http://www.arXiv.org/abs/hep-th/0601021}{{\tt hep-th/0601021}}.

\bibitem{Nielsen:2006th}
H.~B. Nielsen and M.~Ninomiya, ``Compactified time and likely entropy: World
  inside time machine: Closed time-like curve,''
\href{http://www.arXiv.org/abs/hep-th/0601048}{{\tt hep-th/0601048}}.

\bibitem{Kong:2008eu}
D.-X. Kong and K.~Liu, ``{Time-Periodic Solutions of the Einstein's Field
  Equations},''
\href{http://www.arXiv.org/abs/0805.1100}{{\tt 0805.1100}}.

\bibitem{Henneaux:1998ch}
M.~Henneaux, ``{Boundary terms in the AdS/CFT correspondence for spinor
  fields},''
\href{http://www.arXiv.org/abs/hep-th/9902137}{{\tt hep-th/9902137}}.

\bibitem{Csaki:2003dt}
C.~Csaki, C.~Grojean, H.~Murayama, L.~Pilo, and J.~Terning, ``Gauge theories on
  an interval: Unitarity without a higgs,'' {\em Phys. Rev.} {\bf D69} (2004)
  055006,
\href{http://www.arXiv.org/abs/hep-ph/0305237}{{\tt hep-ph/0305237}}.

\bibitem{Matsubara:1955ws}
T.~Matsubara, ``{A New approach to quantum statistical mechanics},'' {\em Prog.
  Theor. Phys.} {\bf 14} (1955)
351--378.

\bibitem{Kaluza:1921tu}
T.~Kaluza, ``On the problem of unity in physics,'' {\em Sitzungsber. Preuss.
  Akad. Wiss. Berlin (Math. Phys. )} {\bf 1921} (1921)
966--972.

\bibitem{Klein:1926tv}
O.~Klein, ``Quantum theory and five-dimensional theory of relativity,'' {\em Z.
  Phys.} {\bf 37} (1926)
895--906.

\bibitem{Dvali:2001gm}
G.~R. Dvali, G.~Gabadadze, M.~Kolanovic, and F.~Nitti, ``{The power of
  brane-induced gravity},'' {\em Phys. Rev.} {\bf D64} (2001) 084004,
\href{http://www.arXiv.org/abs/hep-ph/0102216}{{\tt hep-ph/0102216}}.

\bibitem{Carena:2002me}
M.~Carena, T.~M.~P. Tait, and C.~E.~M. Wagner, ``Branes and orbifolds are
  opaque,'' {\em Acta Phys. Polon.} {\bf B33} (2002) 2355,
\href{http://www.arXiv.org/abs/hep-ph/0207056}{{\tt hep-ph/0207056}}.

\bibitem{Karch:2006pv}
A.~Karch, E.~Katz, D.~T. Son, and M.~A. Stephanov, ``{Linear Confinement and
  AdS/QCD},'' {\em Phys. Rev.} {\bf D74} (2006) 015005,
\href{http://www.arXiv.org/abs/hep-ph/0602229}{{\tt hep-ph/0602229}}.

\bibitem{Broglie:1924}
L.~d. Broglie {\em Phil. Mag.} {\bf 47} (1924) 446.

\bibitem{Broglie:1925}
L.~d. Broglie {\em Ann. Phys} {\bf 3} (1925) 22.

\bibitem{2008FoPh...38..659C}
P.~{Catillon}, N.~{Cue}, M.~J. {Gaillard}, R.~{Genre}, M.~{Gouan{\`e}re}, R.~G.
  {Kirsch}, J.-C. {Poizat}, J.~{Remillieux}, L.~{Roussel}, and M.~{Spighel},
  ``{A Search for the de Broglie Particle Internal Clock by Means of Electron
  Channeling},'' {\em Foundations of Physics} {\bf 38} (July, 2008) 659--664.

\bibitem{'tHooft:2001ar}
G.~'t~Hooft, ``{Determinism in free bosons},'' {\em Int. J. Theor. Phys.} {\bf
  42} (2003) 355--361,
\href{http://www.arXiv.org/abs/hep-th/0104080}{{\tt hep-th/0104080}}.

\bibitem{'tHooft:2001fb}
G.~'t~Hooft, ``{How Does God Play Dice? (Pre-)Determinism at the Planck
  Scale},''
\href{http://www.arXiv.org/abs/hep-th/0104219}{{\tt hep-th/0104219}}.

\bibitem{'tHooft:2001ct}
G.~'t~Hooft, ``{Quantum mechanics and determinism},''
\href{http://www.arXiv.org/abs/hep-th/0105105}{{\tt hep-th/0105105}}.

\bibitem{Arkani-Hamed:2001ca}
N.~Arkani-Hamed, A.~G. Cohen, and H.~Georgi, ``(de)constructing dimensions,''
  {\em Phys. Rev. Lett.} {\bf 86} (2001) 4757--4761,
\href{http://www.arXiv.org/abs/hep-th/0104005}{{\tt hep-th/0104005}}.

\bibitem{Berezhiani:2002et}
Z.~Berezhiani, A.~Gorsky, and I.~I. Kogan, ``{On the deconstruction of time},''
  {\em JETP Lett.} {\bf 75} (2002) 530--533,
\href{http://www.arXiv.org/abs/hep-th/0203016}{{\tt hep-th/0203016}}.

\bibitem{Elze:2002eg}
H.-T. Elze and O.~Schipper, ``{Time without time: A stochastic clock model},''
  {\em Phys. Rev.} {\bf D66} (2002) 044020,
\href{http://www.arXiv.org/abs/gr-qc/0205071}{{\tt gr-qc/0205071}}.

\bibitem{Elze:2003nu}
H.-T. Elze, ``Quantum mechanics emerging from timeless classical dynamics,''
\href{http://www.arXiv.org/abs/quant-ph/0306096}{{\tt quant-ph/0306096}}.

\bibitem{Elze:2003ws}
H.-T. Elze, ``Quantum mechanics and discrete time from 'timeless' classical
  dynamics,'' {\em Lect. Notes Phys.} {\bf 633} (2004) 196,
\href{http://www.arXiv.org/abs/gr-qc/0307014}{{\tt gr-qc/0307014}}.

\bibitem{Elze:2003tb}
H.-T. Elze, ``Emergent discrete time and quantization: Relativistic particle
  with extradimensions,'' {\em Phys. Lett.} {\bf A310} (2003) 110--118,
\href{http://www.arXiv.org/abs/gr-qc/0301109}{{\tt gr-qc/0301109}}.

\bibitem{Nikolic:2006az}
H.~Nikolic, ``{Quantum mechanics: Myths and facts},'' {\em Found. Phys.} {\bf
  37} (2007) 1563--1611,
\href{http://www.arXiv.org/abs/quant-ph/0609163}{{\tt quant-ph/0609163}}.

\bibitem{Dolce:2009ce}
D.~Dolce, ``{Compact Time and Determinism for Bosons (version 4)},'' {\em
  Found. Phys.} (2010) \href{http://www.arXiv.org/abs/0903.3680v4}{{\tt
  0903.3680v4}}. For editorial reasons the present paper (Compact Time and
  Determinism for Bosons: foundations. 0903.3680v5) is limited to the
  foundamental part of the previous version 0903.3680v4 (that is par.1, par.2
  and par.3.2). The remaining parts (that is par.3.1, app.A and app.B) will be
  extended and published in dedicated papers.

\bibitem{1996FoPhL}
R.~{Ferber}, ``{A missing link: What is behind de Broglie's ``periodic
  phenomenon''?},'' {\em Foundations of Physics Letters, Volume 9, Issue 6,
  pp.575-586} {\bf 9} (Dec., 1996) 575--586.

\bibitem{whurt:2005}
A.~{F\"ohlisch}, P.~{Feuler}, F.~{Hennies}, A.~{Fink}, D.~{Manzel},
  D.~{Sanchez-Portal}, P.-M. {Echenique}, and W.~{Wurth}, ``{Direct oservation
  of electron dynamics in the attosecond domain},'' {\em Nature} {\bf 436}
  (July, 2005) 373--376.

\bibitem{Weinberg:1996kr}
S.~Weinberg, ``The quantum theory of fields. vol. 2: Modern applications,''.
  Cambridge, UK: Univ. Pr. (1996) 489 p.

\bibitem{Feynman:1942us}
R.~P. Feynman, ``{The principle of least action in quantum mechanics},''.

\bibitem{Rosenstein:1985wb}
B.~Rosenstein and L.~P. Horwitz, ``{probability current versus charge current
  of a relativistic particle},'' {\em J. Phys.} {\bf A18} (1985)
2115--2121.

\bibitem{Nikolic:2007id}
H.~Nikolic, ``{Is quantum field theory a genuine quantum theory? Foundational
  insights on particles and strings},'' {\em Europhys. Lett.} {\bf 85} (2009)
  20003,
\href{http://www.arXiv.org/abs/0705.3542}{{\tt 0705.3542}}.

\bibitem{Nikolic:2008sn}
H.~Nikolic, ``{Probability in relativistic Bohmian mechanics of particles and
  strings},'' {\em Found. Phys.} {\bf 38} (2008) 869--881,
\href{http://www.arXiv.org/abs/0804.4564}{{\tt 0804.4564}}.

\bibitem{'tHooft:2006sy}
G.~'t~Hooft, ``{The mathematical basis for deterministic quantum mechanics},''
  {\em J. Phys.: Conf. Ser.} {\bf 67} (2007) 15pp,
\href{http://www.arXiv.org/abs/quant-ph/0604008}{{\tt quant-ph/0604008}}.

\bibitem{'tHooft:2007xi}
G.~'t~Hooft, ``{Emergent Quantum Mechanics and Emergent Symmetries},'' {\em AIP
  Conf. Proc.} {\bf 957} (2007) 154--163,
\href{http://www.arXiv.org/abs/0707.4568}{{\tt 0707.4568}}.

\bibitem{Jaffe:2005vp}
R.~L. Jaffe, ``{The Casimir effect and the quantum vacuum},'' {\em Phys. Rev.}
  {\bf D72} (2005) 021301,
\href{http://www.arXiv.org/abs/hep-th/0503158}{{\tt hep-th/0503158}}.

\bibitem{Rayleigh:1905}
L.~Rayleigh, ``{The problem of the Random Walk},'' {\em Nature} {\bf 72} (1905)
  318.

\bibitem{hooft-2009-0}
G.~'t~Hooft, ``Entangled quantum states in a local deterministic theory,''
  2009.

\bibitem{hooft-2009-1}
G.~'t~Hooft, ``Quantum gravity without space-time singularities or horizons,''
  2009.

\bibitem{Elze:2005gv}
H.-T. Elze, ``A quantum field theory as emergent description of constrained
  supersymmetric classical dynamics,'' {\em Brazilian Journal of Physics} {\bf
  35 no 2A+2B} (2005) 205--529,
\href{http://www.arXiv.org/abs/hep-th/0508095}{{\tt hep-th/0508095}}.

\bibitem{Gozzi:1991di}
E.~Gozzi, M.~Reuter, and W.~D. Thacker, ``{Symmetries of the classical path
  integral on a generalized phase space manifold},'' {\em Phys. Rev.} {\bf D46}
  (1992)
757--765.

\bibitem{Gozzi:2006cv}
E.~Gozzi and D.~Mauro, ``{Quantization as a dimensional reduction
  phenomenon},'' {\em AIP Conf. Proc.} {\bf 844} (2006) 158--176,
\href{http://www.arXiv.org/abs/quant-ph/0601209}{{\tt quant-ph/0601209}}.

\bibitem{johnson:227}
S.~C. Johnson and T.~D. Gutierrez, ``Visualizing the phonon wave function,''
  {\em American Journal of Physics} {\bf 70} (2002), no.~3, 227--237.

\bibitem{Satz:2008kb}
H.~Satz, ``{The Thermodynamics of Quarks and Gluons},''
\href{http://www.arXiv.org/abs/0803.1611}{{\tt 0803.1611}}.

\bibitem{Magas:2003yp}
V.~K. Magas, A.~Anderlik, C.~Anderlik, and L.~P. Csernai, ``{Non-equilibrated
  post freeze out distributions},'' {\em Eur. Phys. J.} {\bf C30} (2003)
  255--261,
\href{http://www.arXiv.org/abs/nucl-th/0307017}{{\tt nucl-th/0307017}}.

\bibitem{Pomarol:2000hp}
A.~Pomarol, ``{Grand unified theories without the desert},'' {\em Phys. Rev.
  Lett.} {\bf 85} (2000) 4004--4007,
\href{http://www.arXiv.org/abs/hep-ph/0005293}{{\tt hep-ph/0005293}}.

\bibitem{ArkaniHamed:2000ds}
N.~Arkani-Hamed, M.~Porrati, and L.~Randall, ``{Holography and
  phenomenology},'' {\em JHEP} {\bf 08} (2001) 017,
\href{http://www.arXiv.org/abs/hep-th/0012148}{{\tt hep-th/0012148}}.

\bibitem{Maldacena:1997re}
J.~M. Maldacena, ``{The large N limit of superconformal field theories and
  supergravity},'' {\em Adv. Theor. Math. Phys.} {\bf 2} (1998) 231--252,
\href{http://www.arXiv.org/abs/hep-th/9711200}{{\tt hep-th/9711200}}.

\bibitem{Witten:1998qj}
E.~Witten, ``{Anti-de Sitter space, thermal phase transition, and confinement
  in gauge theories},'' {\em Adv. Theor. Math. Phys.} {\bf 2} (1998) 505--532,
\href{http://www.arXiv.org/abs/hep-th/9803131}{{\tt hep-th/9803131}}.

\end{thebibliography}\endgroup
\end{document}